\documentclass[fleqn,usenatbib]{mnras}
\usepackage{newtxtext,newtxmath}
\usepackage[T1]{fontenc}

\usepackage{lscape}

\DeclareRobustCommand{\VAN}[3]{#2}
\let\VANthebibliography\thebibliography
\def\thebibliography{\DeclareRobustCommand{\VAN}[3]{##3}\VANthebibliography}

\makeatletter
\newlength{\abovecaptionskip}%
\makeatother

\usepackage{graphicx}	% Including figure files
\usepackage{amsmath}	% Advanced maths commands
\usepackage{threeparttable}
\usepackage{adjustbox}
\usepackage{color}
\usepackage{booktabs}

\newcommand{\masyr}{ \ {\rm{mas \ yr^{-1}}}\>}
\newcommand{\mas}{ \ {\rm{mas}}\>}

\newcommand{\kms}{ \ {\rm{km \ s^{-1}}}\>}

\newcommand{\foo}[1]{}

\newcommand{\degree}{\degr}

\newcommand{\parallax}{\varpi}
\newcommand{\vlos}{v_{\mathrm{los}}}

\newcommand{\hyperfootnote}[1][]{\def\ArgI\hyperfootnoteRelay}
\newcommand\hyperfootnoteRelay[2][]{\href{#1#2}{\ArgI}\footnote{\href{#1#2}{#2}}}
\title[Internal rotation with EDR3]{Internal rotation of Milky Way dwarf spheroidal satellites with \textit{Gaia} Early Data Release 3}

\author[A. M. Martínez-García et al.]{
Alberto Manuel Martínez-García,$^{1,2}$\thanks{E-mail: ammtnez@iac.es}
Andrés del Pino,$^{3}$
Antonio Aparicio,$^{1,2}$
\newauthor
Roeland P. van der Marel,$^{3,4}$
and Laura L. Watkins$^{5}$ %ORCID: 0000-0002-1343-134X
\\
$^{1}$Instituto de Astrofísica de Canarias, Calle Vía Láctea S/N, E-38205 La Laguna, Tenerife, Spain\\
$^{2}$Universidad de La Laguna, Dpto. Astrofísica, Avda. Astrofísico Fco. Sánchez S/N, E-38206 La Laguna, Tenerife, Spain\\
$^{3}$Space Telescope Science Institute, 3700 San Martin Drive, Baltimore, MD 21218, USA \\
$^{4}$Center for Astrophysical Sciences, Department of Physics \& Astronomy, Johns Hopkins University, Baltimore, MD 21218, USA\\
$^{5}$AURA for the European Space Agency (ESA), ESA Office, Space Telescope Science Institute, 3700 San Martin Drive, Baltimore, MD 21218, USA 
}

\date{Accepted: 2021 May 26; Revised: 2021 May 26; Received: 2021 April 1}

\pubyear{2021}

\begin{document}
\label{firstpage}
\pagerange{\pageref{firstpage}--\pageref{lastpage}}
\maketitle

% Abstract of the paper
\begin{abstract}
We present an analysis of the kinematics of 14 satellites of the Milky Way (MW). We use proper motions (PMs) from the \textit{Gaia} Early Data Release 3 (EDR3) and line-of-sight velocities ($\vlos$) available in the literature to derive the systemic 3D motion of these systems. For six of them, namely the Carina, Draco, Fornax, Sculptor, Sextans, and Ursa Minor dwarf spheroidal galaxies (dSph), we study the internal kinematics projecting the stellar PMs into radial, $V_R$ (expansion/contraction), and tangential, $V_T$ (rotation), velocity components with respect to the centre of mass. We find significant rotation in the Carina ($|V_T| = 9.6 \pm 4.5 \kms$), Fornax ($|V_T| = 2.8 \pm 1.3 \kms$), and Sculptor ($|V_T| = 3.0 \pm 1.0 \kms$) dSphs. Besides the Sagittarius dSph, these are the first measurements of internal rotation in the plane of the sky in the MW's classical dSphs. All galaxies except Carina show $|V_T| / \sigma_v < 1$. We find that slower rotators tend to show, on average, larger sky-projected ellipticity (as expected for a sample with random viewing angles) and are located at smaller Galactocentric distances (as expected for tidal stirring scenarios in which rotation is transformed into random motions as satellites sink into the parent halo). However, these trends are small  and not statistically significant, indicating that rotation has not played a dominant role in shaping  the 3D structure of these galaxies. Either tidal stirring had a weak impact on the evolution of  these systems or it perturbed them with similar efficiency regardless of their current Galactocentric distance.
\end{abstract}

% Select between one and six entries from the list of approved keywords.
% Don't make up new ones.

%%%%%%%%%%%%%%%%%%%%%%%%%%%%%%%%%%%%%%%%%%%%%%%%%%

%%%%%%%%%%%%%%%%% BODY OF PAPER %%%%%%%%%%%%%%%%%%

\section{Introduction}

The formation and evolution of galaxies remain a crucial question yet to be solved. Dwarf galaxies play a central role in the context of the $\Lambda$ Cold Dark Matter  ($\Lambda$CDM) model since they are thought to be the first galaxies formed and to have given rise to larger structures through mergers \citep{WhiteRees1978, Blumenthal1984, Dekel1986, NavarroFrenkWhite1995}. Therefore, dwarf galaxies observed nowadays could be the remaining systems that have not yet merged into larger galaxies and could contain key information about the primeval Universe.

Dwarf galaxies show a wide variety in their morphologies, star formation histories, and masses that have raised questions about the possible mechanisms involved in their formation and evolution. A first major taxonomy of dwarfs divides them into dwarf irregulars (dIrrs) and dwarf ellipticals (dEs), of which the fainter subset are called dwarf spheroidals (dSphs). dIrrs are gas-rich, star-forming galaxies that may be rotation-supported and discy to some extent (\citealt{Mateo1998, McConnachie2012, Kirby2014, Wheeler2017}), while dSphs are pressure-supported, gas-poor systems with spheroidal morphology (\citealt{McConnachie2012, Wheeler2017}). Ultra faint dwarfs (UFDs) are a further sub-type with characteristics in common with dSphs except that they 
have lower luminosities and surface brigthnesses (\citealt{Simon2019}, and references therein). For simplicity, we will use sometimes the common term {\it early-type} to refer to both dSphs and UFDs together, and {\it late-type} for dIrrs.

Most satellites of the Milky Way (MW) are dSph or UFD galaxies. They are highly dark-matter dominated with mass-to-light ratios $(M/L) \geq 100 {\rm\, M}_\odot/{\rm L}_\odot$ in many cases \citep{Mateo1998, Kleyna2005, McConnachie2012} and have low optical central surface brightness ($\mu_{V,0} \sim 26-30$ mag arcsec$^{-2}$,  \citealp{Gallagher1994,McConnachie2012}) and low luminosities ($-13 \leq M_V \leq -4$, \citealt{Mateo1998, Belokurov2007}). They are extremely gas-poor \citep{Gallagher1994} and host mainly old stellar populations \citep[$\gtrsim$ 10 Gyr,][]{Gallagher1994, Aparicio2001, Carrera2002, Tolstoy2004, Battaglia2006, Monelli2010a, Monelli2010b, deBoer2011, Hidalgo2013, Bettinelli2018, Bettinelli2019}. Some of them have relatively complex stellar populations, showing evidence of an intermediate-age stellar population, in particular, Sculptor, Fornax, Carina, and Sextans \citep{Kleyna2004, Tolstoy2004,  Battaglia2006, Koch2008, DelPino2013, Bettinelli2018, Bettinelli2019}.

From a cosmological point of view, early-type dwarfs are likely to be examples of the evolved state of late-type dwarfs, transformed into spheroids as a result of tidal stirring, ram pressure and other environmental effects \citep[][and references therein]{Mayer2010}. These mechanisms had been proposed to explain the morphology--density relation \citep{Mateo1998} observed among Local Group dwarfs: early-type galaxies tend to be located close to the largest galaxies of the group, whereas dwarf irregulars are usually found in the outskirts of the group. If this scenario is correct, some residual rotation signal should be present in the observed dwarfs around the MW \citep{Kazantzidis2011, Lokas2015}.

A great observational effort has been made over the last few decades to measure line-of-sight velocities ($\vlos$) of stars in MW satellites (see \citealt{Walker2012} or \citealt{Fritz2018} for a comprehensive list of kinematic studies). Most dSphs do not show obvious signs of rotation \citep{Wilkinson2004, Munoz2005, Koch2007a, Koch2007b, Walker2009}. Thus far, only Sagittarius, Fornax, Sculptor, and Sextans have been found to exhibit signs of rotation related to $\vlos$ gradients \citep{Battaglia2008b, Battaglia2011, Amorisco2012, Zhu2016, delPino2017, delPino2021}, and very recently some hints for prolate rotation have been found in Ursa Minor \citep{Pace2020}. However, it is unclear whether these gradients are an intrinsic rotation signal or rather the projection in the sky of the tangential component of the velocity \citep[see, for example, ][]{Walker2008, Strigari2010}. Therefore, early-type dwarfs are generally considered to be pressure-supported systems with little or no rotation.

The unprecedented astrometric capabilities of the {\it Gaia} mission (\citealt{GaiaCollaboration2016}), have made it possible to measure the tangential component of velocity, or proper motion (PM), of thousands of stars in dwarfs satellites of the MW (\citealt{Helmi2018}).
In this work, we use the {\it Gaia} Early Data Release 3 (EDR3, \citealt{GaiaEDR3}) to derive the systemic PM of 14 galaxies and to study, for the first time, the internal kinematics of six of them, namely Carina, Draco, Fornax, Sculptor, Sextans, and Ursa Minor, using the PMs of their individual stars. We also study possible relations between their internal kinematics, morphology, and their location with respect to the MW.

The paper is organized as follows. In Section~\ref{sec:data}, we introduce the data set, the membership selection procedure, and the methods. In Section~\ref{sec:Samples_of_galaxies}, we introduce the galaxy samples. In Section~\ref{sec:discus}, we present and discuss the results. Finally, in Section~\ref{sec:concl} we summarize the conclusions of this work.

\section{Data and methods}
\label{sec:data}

%\begin{landscape}
\begingroup
\begin{table*}
    \caption{Fundamental parameters of the galaxies of \textit{sample 1}, ordered by Galactocentric distance. Columns: (1) galaxy name, (2) right ascension, (3) declination, (4) systemic velocity, (5) heliocentric distance, (6) half-light radius, (7) tidal radius, (8) central surface brightness, (9) mean surface brightness within the isophote defined by the half-light radius, and (10)  references: (1) \citet{McConnachie2020a}; (2) \citet{McConnachie2012}; (3) \citet{Helmi2018}; (4) \citet{Roderick2016}; (5) \citet{Okamoto2012}; and (6) \citet{Irwin1995}, and references therein.}

	\setlength{\tabcolsep}{3.3pt}
	\begin{tabular}{l|c|c|c|c|c|c|c|c|c|c|c|c}  
	    \hline
		Galaxy & RA J2000& DEC  J2000& $\vlos$  &$D_0$ & $r_{h}$ & $r_{t}$ & $\mu_V$ & $\mu_{\mathrm{eff}}$ & Ref \\
		 & (deg) & (deg) & (km s$^{-1}$) & (kpc)  & (arcmin) & (arcmin) &  (mag arcsec$^{-2}$) & (mag arcsec$^{-2}$) & \\
		\hline
		\hline % $\pm$ &
		Reticulum II     & 53.9254  & --54.0492 & 64.7$^{+1.3}_{-0.8}$ & 30.2 $\pm$ 2.8 & 5.59 $\pm$ 0.21 & -- & -- & -- & (1)\\
		Bootes I         & 210.0250  & 14.5000   & 99.0 $\pm$ 2.1   & 66.4 $\pm$ 2.4 &   12.50 $\pm$ 0.30 & 31.6 $\pm$ 4.0 & 27.5 $\pm$ 0.3 & 28.7& (1), (2), (3), (4), (5)\\
		Sagittatius II   & 298.1688 & --22.0681 & --177.3 $\pm$ 1.2 & 73.1 $\pm$ 1.0 & 1.70$\pm$ 0.05 & -- & -- & -- &  (1) \\
		Draco            & 260.0517 & 57.9153  & --291.0 $\pm$ 0.1 & 75.9 $\pm$ 5.9 &  9.93 $\pm$ 0.09 & 28.3 $\pm$ 2.4 & 25.0 $\pm$ 0.2  & 26.1 & (1), (2), (3), (6)\\
		Ursa Minor       & 227.2854 & 67.2225  & --246.9 $\pm$ 0.1 & 75.9 $\pm$ 3.5 &  17.32 $\pm$ 0.11 &  50.6 $\pm$ 3.6 & 26.0 $\pm$ 0.5 & 25.2 & (1), (2), (3), (6)\\
		Sculptor         & 15.0392  & --33.7092 & 111.4 $\pm$ 0.1  & 85.9 $\pm$ 5.5 & 12.33 $\pm$ 0.05 & 76.5 $\pm$ 5.0 & 23.5 $\pm$ 0.5 &  24.3 & (1), (2), (3), (6)\\
		Sextans          & 153.2625 & --1.6147  & 224.2 $\pm$ 0.1  & 85.9 $\pm$ 4.0 &    27.80 $\pm$ 1.20 & 160.0 $\pm$ 50.0 & 27.1 $\pm$ 0.5 & 28.0 & (1), (2), (3), (6)\\
		Ursa Major I     & 158.7200 & 51.9200 & --55.3 $\pm$ 1.4 & 96.8 $\pm$ 4.5 &   8.34  $\pm$  0.34 & -- & 27.7 $\pm$ 0.5 & 28.8 & (1), (2) \\
		Carina           & 100.4029 & --50.9661 & 222.9 $\pm$ 0.1  & 105.2 $\pm$ 6.3 &   11.43 $\pm$ 0.12 & 28.8 $\pm$ 3.6 & 25.5 $\pm$ 0.5  & 26.0 & (1), (2), (3), (6)\\
		Crater II        &  177.3100 & --18.4131 & 87.5 $\pm$ 0.4 & 117.5 $\pm$ 1.1 & 31.20 $\pm$ 2.50 & --& -- & -- & (1) \\
		Fornax           & 39.9971  & --34.4492 &   55.3 $\pm$ 0.1 & 147.2 $\pm$ 12.2 &  18.40 $\pm$ 0.20 & 71.1 $\pm$ 4.0  & 23.3 $\pm$ 0.3 & 24.0 & (1), (2), (3), (6)\\
		C. Venatici I & 202.0146 & 33.5558 & 30.9 $\pm$ 0.6 & 217.8 $\pm$ 10.0 &  8.90 $\pm$ 0.40  & -- & 27.1 $\pm$ 0.2 & 28.2 & (1), (2) \\
		Leo II           & 168.3700 & 22.1517  & 78.0 $\pm$ 0.1   & 233.3 $\pm$ 14.0 &  2.48  $\pm$ 0.03  &  8.7 $\pm$ 0.9 & 24.2 $\pm$ 0.3 & 24.8 &  (1), (2), (3), (6)\\
		Leo I            & 152.1171 & 12.3064  & 282.5 $\pm$ 0.1  & 253.5 $\pm$ 15.2 &  3.30 $\pm$ 0.03& 12.6 $\pm$ 1.5  & 22.6 $\pm$ 0.3 & 23.3  & (1), (2), (3), (6)\\
		\hline
	\end{tabular}
	\label{tab:properties}
\end{table*}
\endgroup
%\end{landscape}

To study the kinematics of a dwarf galaxy, we need to select its stellar members and to derive their velocities with respect to a co-moving reference frame. In this section, we provide details of the stellar membership selection process and the coordinate transformation. This will allow us to study the galaxies contained in the three galaxy samples discussed in Section~\ref{sec:Samples_of_galaxies}. Readers interested mostly in how these samples are selected may wish to skip directly to that section.

\subsection{Stellar membership selection}

\subsubsection{Quality cuts}\label{subsec:cuts}

To select member stars in each dwarf, we use \textsc{GetGaia}\footnote{https://github.com/AndresdPM/GetGaia} \citep{delPino2021}, which performs the tasks of downloading the data, selecting members, and screening out poorly measured stars through a series of nested criteria. We proceed as follows.

For each galaxy, we download all the stars from the \textit{Gaia} \texttt{gaia\_source} table within a circular region of radius $r\leq 1.5r_t$, where $r_t$ is the tidal radius (taken from the literature, 
\citealt{Irwin1995, Roderick2016}). For galaxies for which $r_t$ is not available, the criterion $r\leq 6r_h$ is used instead, where $r_h$ is the half-light radius (\citealt{McConnachie2020a}).
Only stars with available colours and astrometric solution are considered. We impose parallax, $\parallax$, and PM cuts to select stars compatible with the bulk and the internal dynamical properties of each galaxy. In particular, we select stars whose PMs are in the range $\mu_0 - 2.5 \masyr \leq \mu \leq \mu_0 + 2.5 \masyr$, where $\mu_0$ is the PM in the RA and Dec. directions from \citet{McConnachie2020b}. For $\parallax$, we require the stars to be in the range of $-2.5 \mas \leq \parallax \leq 1.5 \mas$. Following \textit{Gaia} EDR3 verification papers, we consider only sources with:
\begin{itemize}
    \item $\text{RUWE} \leq 1.4$
    \item $\mbox{\tt ipd\_gof\_harmonic\_amplitude} \leq 0.2$
    \item $\mbox{\tt visibility\_periods\_used} \geq 9$
    \item $\mbox{\tt astrometric\_excess\_noise\_sig} \leq 2$
    \item $|C^*| \leq 3\sigma(C^*)$
\end{itemize}
where $\text{RUWE}$ is the renormalized unit weight error and $C^*$ is the corrected flux excess factor defined in \citet{Riello2020}. We then correct parallaxes using the prescription suggested in \cite{Lindegren2020}.

A further selection criterion is related to astrometric errors. We account for those being 1.05 and 1.22 times larger than the ones listed in the {\tt gaia\_source} table for 5-parameter and 6-parameter solutions, respectively \citep{Fabricius2020}. As for small-scale systematic errors, still present in EDR3, we simulate them for each star following the prescription explained in \citet{VanderMarel2019} and \citet{delPino2021}. We use a smaller amplitude, of 56 $\mu$as yr$^{-1}$, which provides a root-mean-square consistent with the one reported by \citet{Lindegren2020} at scales of $\sim 1\degree$.

We select stars based on their astrometric and photometric errors, keeping only stars with errors below the typical nominal ones in the EDR3 at $G=21$. Namely, only stars with errors smaller than $1 \masyr$ in PM, $0.7 \mas$ in $\parallax$, 0.01 mag in $\mbox{\tt phot\_g\_mean\_mag}$ and 0.1 mag in $\mbox{\tt phot\_bp\_mean\_mag}$ and $\mbox{\tt phot\_rp\_mean\_mag}$ are conserved.

A last cleaning of the sample is performed by screening out sources with magnitudes \linebreak 
$\mbox{\tt phot\_bp\_mean\_mag} > 20.5$ and $\mbox{\tt phot\_rp\_mean\_mag} > 20.5$ to avoid the observed bias in color at faint magnitudes \citep{Fabricius2020, Riello2020}.

\subsubsection{Membership algorithms}\label{subsec:data}

{\sc GetGaia}'s membership selection is based on the serial application of three different steps: likelihood of membership based on position of the star in the sky, CMD, and PM-parallax ({\it step 1}), a refinement of the PM-parallax selection through a Gaussian clustering ({\it step 2}), and further removal of PM and parallax contaminants based on a Voronoi tessellation ({\it step 3}).

In {\it step 1}, we consider that stars in the vicinity of a dwarf will be a mix of true dwarf stars and contaminating field stars. We use a sample of stars away from the dwarf to define the properties of the field sample and then use these to infer the properties of dwarf stars in the region of interest. We do this separately in 2D sky coordinates space, 2D colour-magnitude space, and 3D PM-parallax space. This procedure is explained in detail in Watkins et al. (in preparation), but we include a brief summary here.

For each star $i$ in the sample of interest, we calculate the probability of finding another star in the sample on the sky, $\alpha$-$\delta$ $P^\mathrm{Sample}_{\mathrm{sky},i}$ through a Multi-Gaussian Expansion parametrization of the galaxy surface brightness with the fitting algorithm of \citet{Cappellari2002}. This probability is combined with those of finding the same star in colour-magnitude $P^\mathrm{Sample}_{\mathrm{CMD},i}$ or PM-parallax $P^\mathrm{Sample}_{\mathrm{PMP},i}$. We then calculate the probability of finding a star in the field-only sample nearby in $\alpha$-$\delta$ $P^\mathrm{Field}_{\mathrm{sky},i}$, colour-magnitude  $P^\mathrm{Field}_{\mathrm{CMD},i}$ or PM-parallax $P^\mathrm{Field}_{\mathrm{PMP}},i$. Finally, we assume that the field stars have a constant density across both samples and so can use the relative areas spanned by the sample of interest and the field-only sample to estimate the probability of finding a field star in the region of interest at the position of the given star. That is, the probability of finding a dwarf star is then the total probability of finding a star in the sample of interest, less the probability of finding a field star:
\begin{equation}
    P^\mathrm{Dwarf}_i =  P^\mathrm{Sample}_i - f P^\mathrm{Field}_i ,
\end{equation}
where $f = A_\mathrm{Sample}/A_\mathrm{Field}$ is the correction factor based on the areas $A$.

Only stars with a probability value above a certain value are considered as possible members and pass to the next selection step. The values used for selection were chosen individually for each galaxy. Specifically, we consider the distribution in $\alpha$-$\delta$, colour-magnitude, and PM-parallax of the possible member and non-member stars, aiming to obtain a uniform density in these spaces for the non-members and no presence of members in areas where such stars are not expected to be present e.g. bright blue stars in the CMD for a galaxy that is composed solely by old stars.

In {\it step 2}, stars that pass {\it step 1} are then clustered and selected in PM-$\parallax$ space. A 3D Gaussian model is fitted to the data in the PM-$\parallax$ space and stars are scored based on their logarithmic likelihood of belonging to the Gaussian distribution. The Gaussian model fitting is performed iteratively. As a first guess for the parameters, we used the PMs from \citet{McConnachie2020b} and $\parallax_0 = 0$. Stars are then scored based on their logarithmic likelihood of belonging to the obtained Gaussian distribution. Stars not fulfilling the condition $\ell_\star \geq \tilde{\ell} - 3 \sigma(\ell)$ are rejected, where $\ell_\star$ is the star score; $\tilde{\ell}$ is the median of all the scores; and $\sigma(\ell)$ is the standard deviation of the score distribution. A new Gaussian fit is performed and the process is repeated until convergence.

This procedure guarantees that stars not compatible with the bulk PMs or parallax are rejected, yet some MW stars may lie inside the distribution of selected stars in the PMs-parallax space and thus be selected as possible members. Most of these contaminants, however, will show PMs and parallaxes not consistent with those from their surrounding neighbour stars, as they follow different trends in such quantities across the sky. In {\it step 3}, stars that pass the {\it step 2} are further filtered to screen out most of these contaminants. To do this, we divide the sky in several Voronoi cells (\citealt{Capellari2003, delPino2017}). Voronoi cells are constructed using the sky stellar coordinates so that each contains approximately the same number of stars. Specifically, in this case, we impose $max(9, \sqrt{N_*})$ stars per cell. The stars are then 3-sigma clipped based on the PMs and parallax within each cell. The tessellation is repeated until no more stars are rejected.

\subsection{Systemic motion derivation}
\label{subsec:CM}
The motion of the centre of mass (CM) of the galaxies must be subtracted from their stars before the study of the internal kinematics. We calculate the 3D velocity vector of the CM using the systemic $\vlos$ values from the literature (\citealt{McConnachie2020a}) and PMs of the galaxy, which were obtained from our member samples.
We compute the systemic PMs as the weighted average of the PMs of individual stars for each galaxy. 

The accurate estimation of the systemic motion of the galaxies requires the correction of potential systematic errors in the astrometric zero-point. To investigate and correct such errors, we use quasars as reference for zero-point astrometric measurements. The zero-points for each galaxy were computed as follows.

We first extract all quasars located within 3 degrees around the central coordinates of each galaxy from the table \texttt{agn\_cross\_id}, provided within the \textit{Gaia} EDR3. This guarantees a sizeable number of quasars ($\sim$ 1000) in the vicinity of the galaxies and yields errors for the correction of the order of those derived for the systemic PMs.
We then impose several constraints to build a solid sample of quasars. Only quasars belonging to the 5-parameter solution are considered, since it has been reported to be more precise than the 6-parameter one (\citealt{Fabricius2020, GaiaEDR3,  Lindegren2020}). We further restrict the selection to effective wavenumbers between 1.24 and 1.72 $\mu$m$^{-1}$ (\citealt{Lindegren2020b}) and screen out possible non-single objects using RUWE < 1.4, \texttt{ipd\_frac\_multi\_peak} $\leqslant$ 2,  and \texttt{ipd\_gof\_harmonic\_amplitude} $\leqslant$ 0.1 (\citealt{Fabricius2020}). Lastly, we filter potential outliers in PM at 5$\sigma$. The astrometric PM zero-point and its associated error are then computed as the weighted average of the PMs of the quasars sample for each galaxy. Finally the zero-point is subtracted from the systemic PM of the corresponding galaxy.

\subsection{Internal kinematics derivation}
\label{sec:v1v2v3}

To study the galaxies' internal kinematics, we introduce a co-moving reference frame centred on each galaxy. The formulation is identical to that presented in \citet{VanderMarel2001} and \citet{VanderMarel2002} and further explained in \citet{delPino2021}. In short, the motion of the CM (Section~\ref{subsec:CM}) is subtracted from the sky-projected velocity field, and individual stellar velocities are decomposed into three orthogonal components:
\begin{equation}\label{eq:v1v2v3}
v_{S,i} \equiv \frac{dD_i}{dt},\\ v_{R,i} \equiv D_i\frac{d\rho_i}{dt},\\ v_{T,i} \equiv D_i\sin{\rho_i}\frac{d\phi_i}{dt} ,
\end{equation}  
where $D_i$ is the heliocentric distance in kpc, $\rho_i$ is the angular distance to the CM, and $\phi_i$ is the position angle (measured North to East) of the $i$-th star. Therefore, $v_{S,i}$ is the line-of-sight component of the velocity, and $v_{R,i}$, $v_{T,i}$ are the radial and tangential components of the velocity in the plane of the sky, being the three components referred with respect to the CM.

For each galaxy, we assume that all stars are located at the same distance, $D_0$, and have the CM $\vlos$ (\citealt{McConnachie2020a}). While this is not exact, the assumption has a negligible impact on $v_{S,i}$, $v_{R,i}$, and $v_{T,i}$, with differences well below the observational errors. 

We propagate the corrected random errors (see section~\ref{subsec:cuts}) of all the variables used during the computation of $v_{S,i}$, $v_{R,i}$, and $v_{T,i}$ through a Monte-Carlo scheme. A total of $10^4$ realizations are performed, randomly sampling each variable from a normal distribution centred on its corresponding nominal value and perturbing them following simulated random small-scale systematic patterns (as described in~\ref{subsec:cuts}). Therefore, the total error for the velocity components $v_{S,i}$, $v_{R,i}$, and $v_{T,i}$ is given as the standard deviation of all the realizations for each star [henceforth $\sigma(v_{S,i})$, $\sigma(v_{R,i})$, and $\sigma(v_{T,i})$, respectively]. Finally, we determine the mean radial ($V_R$) and tangential ($V_T$) components of the velocity of each galaxy as the weighted average of the velocity components of their stars:
\begin{equation}
\label{eq:weightedaverages}
    V_j =\frac{\sum_{i=1}^{n} \frac{v_{j,i}}{\sigma( v_{j,i})^2}}{\sum_{i=1}^{n} \frac{1}{\sigma(v_{j,i})^2}}, \\
    \sigma(V_j) = \left(\frac{1}{\sum_{i=1}^{n} \frac{1}{\sigma( v_{j,i})^2}}\frac{n}{n-1}\right)^{1/2}.
\end{equation}
where $j = R, T$. Note that we do not provide the calculation of $V_S$ since we do not use $\vlos$ for individual stars.

\section{Samples of Galaxies}\label{sec:Samples_of_galaxies}

To study the kinematics of a dwarf galaxy, we need to ensure that we have both enough stars measured and sufficient precision. This limits the sample of galaxies that can be analysed, depending on the information to be obtained. We started by considering all the 58 galaxies listed in \citet{McConnachie2020b} and selecting their member stars from the Gaia EDR3 catalogue applying the methodology described in Section~\ref{sec:data}. We created three successive sub-samples depending on the amount and precision of the data obtained for each galaxy, with the purpose of studying them with successive levels of depth and detail.

\begin{figure*}
    \centering
    \includegraphics[width=1.0\textwidth, scale=0.22]{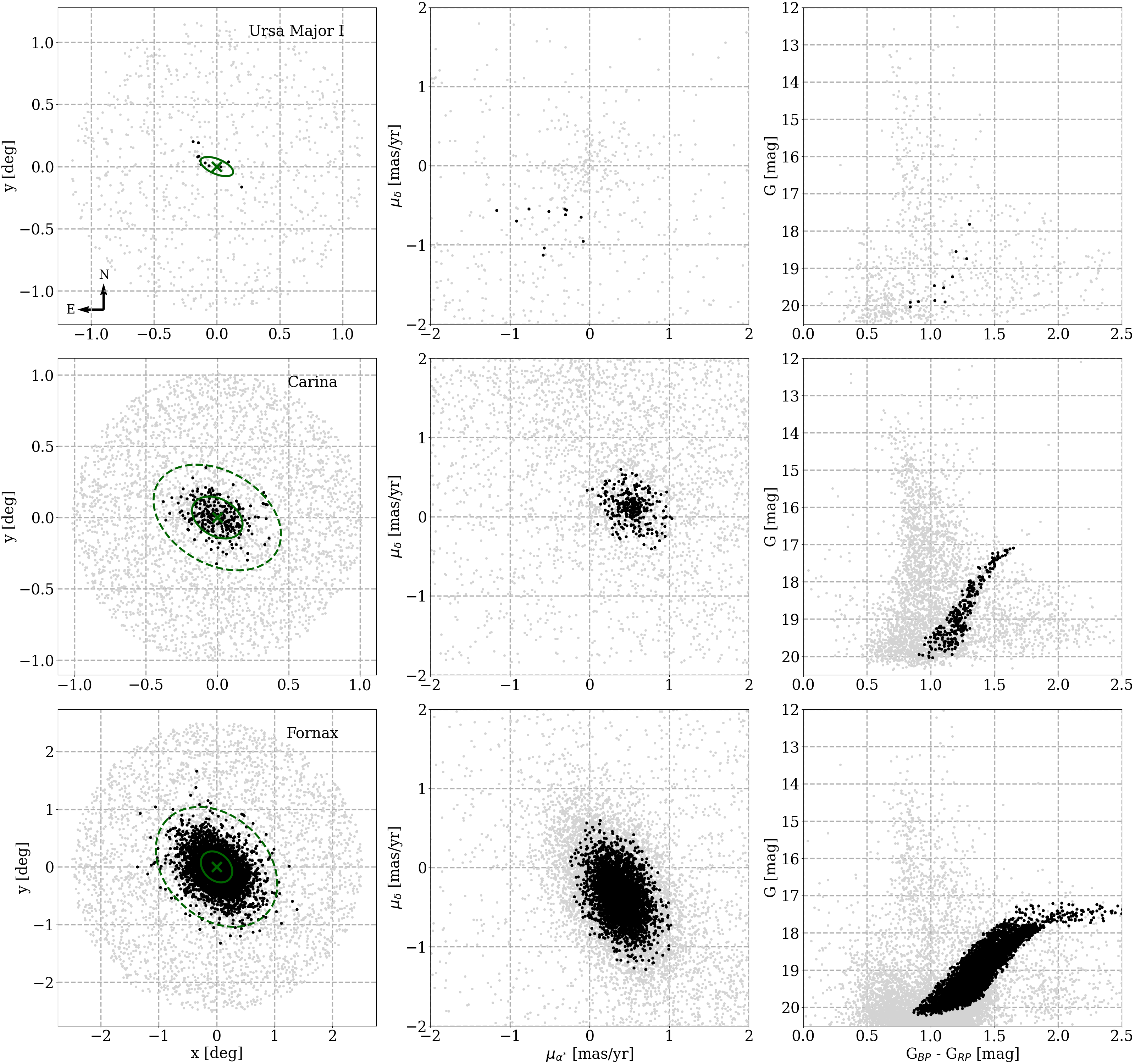}
    \caption{Selection of members from \textit{Gaia} EDR3 for three galaxies from \textit{sample 1}: Ursa Major I, Carina, and Fornax. Left: the spatial distribution of the stars in the sky. Middle: the distribution of the stars in the PM space. Right: the CMDs of the stars. In all panels, the black points represent the selected member stars while grey points represent the sources extracted from \textit{Gaia} EDR3 that were rejected by the selection procedure (see Section~\ref{subsec:data}). In the leftmost panels, the centre of the galaxies is located at $(0,0)$, and marked by a cross. Solid and dashed ellipses represent, respectively, the half-light and tidal radii of the galaxies.
    }
    \label{fig:member_selection}
\end{figure*}

\textit{Sample 1} contains galaxies for which we were able to derive statistically reliable systemic PMs, a total of 14 galaxies. These each contain at least 10 identified member stars, although most of them comprise far more, the median value is $\sim150$ stars. The galaxies in this sample, and their fundamental parameters, can be found in  Table~\ref{tab:properties}. Their derived systemic PMs can be found in Table~\ref{tab:resultadosPM}. In Fig.~\ref{fig:member_selection}, we show the final selection of sources for three galaxies; the one comprising the highest number of members (Fornax), an intermediate case (Carina), and the one with the fewer members (Ursa Major I). The different panels show the distribution in the sky, the PM, and the color-magnitude (CMDs) spaces for the selected and rejected stars. 

{\sc GetGaia} seems to provide a consistent selection of member stars. The vast majority of member stars lie within the corresponding tidal radii ($\sim3 r_h$ in the case of Ursa Major I, for which no tidal radius is available); they form Gaussian-shaped clusters in PM space; and they are located in the Red and Asymptotic Giant Branches (RGB and AGB), as expected in galaxies hosting mainly old and intermediate-age stellar populations.

However, despite the absence of obvious outliers in these samples, some contaminants may have passed through our screening process. To independently assess the contamination rate in our samples, we made use of external catalogues with $\vlos$ from spectroscopy, where available. Specifically, we cross-matched our lists of member stars selected by {\sc GetGaia} with spectroscopic lists of giant stars in Carina, Draco, Fornax, Sculptor, and Sextans from \citet{Walker2009}, and \citet{ Walker2015}. We then calculated the percentage of false positives among stars found in the two lists (false members) by selecting members in PM-$\vlos$ space. Our experiments yielded a median contamination rate of 2.0\%, where Carina is the worst case with 6.3\% of the sample polluted by MW stars, and Fornax the best one with just 1.2\% contamination. We expect these contaminants to have a small impact in the results of internal kinematics, since they represent a small fraction in the samples and due to the fact that they have endured the selection procedure; therefore, they show indistinguishable kinematics from actual member stars as a bulk and within Voronoi cells.

The study of the internal kinematics, on the other hand, requires more restrictive selection criteria. To obtain meaningful results, the errors in the derived internal velocity components should be smaller than the velocity dispersion of the system. Therefore, for the study of the mean sky-projected internal kinematics, we further restrict our selection of galaxies to only those with errors in the mean velocity components, $V_T$ and $V_R$, smaller than the line-of-sight velocity dispersion, $\sigma_v$. This criterion leaves a list of six dSphs, namely the Carina, Draco, Fornax, Sculptor, Sextans, and Ursa Minor dSphs, whose numbers of identified members range from 244 to 5488 (see Table~\ref{tab:resultadosPM}), and that we call \textit{sample 2}. Their mean internal velocity components can be found in Table~\ref{tab:resultadosv}.

Lastly, \textit{sample 3} includes only galaxies whose number of identified members are large enough to study their spatially resolved internal kinematics. We required a minimum of 500 stars per galaxy, which resulted in a list of three galaxies, namely the Fornax, Sculptor, and Ursa Minor dSphs.

\section{Results and discussion}
\label{sec:discus}

\subsection{Systemic 3D motions}
\label{subsec:PMs}

Systemic PMs of the galaxies are needed to derive their systemic 3D motion. These will be later used to subtract the CM motion from the observed stellar velocities when we study the internal kinematics. We derived the systemic PMs following the procedure described in Section~\ref{subsec:CM}. The results for \textit{sample 1}, i.e. galaxies for which we obtain at least 10 member stars, can be found in Table~\ref{tab:resultadosPM}. The table contains the systemic PMs with and without astrometric zero-point correction. The median values of these corrections are 0.007 and 0.004 mas yr$^{-1}$ for the PMs in the RA and Dec. directions respectively. 
Error intervals of Table~\ref{tab:resultadosPM} also include small-scale systematic effects.

%\begin{landscape}
\begin{table*}
\caption{PM results for the \textit{sample 1}, formed by the 14 galaxies for which we were able to identify enough member stars and reach sufficient precision to perform a PM analysis. Columns: (1) galaxy name, (2) Galactocentric distance, (3) number of stars in the final sample, (4) number of quasars used for computing the zero-points, (5) PM in Right Ascension, (6) PM in Declination, (7) PM in Right Ascension corrected for astrometric zero-point, and (8) PM in Declination corrected for astrometric zero-point. For the PMs, we provide random errors and total errors (within parenthesis).}
	\begin{tabular}[b]{l|r|r|r|r|r|r|r} 
	    \hline
		Galaxy & \multicolumn{1}{c}{$d_{\mathrm{GC}}$} & \multicolumn{1}{c}{$n_{*}$} & \multicolumn{1}{c}{$n_{\mathrm{Q}}$} & \multicolumn{1}{c}{$\mu_{\alpha^{*}}$} &  \multicolumn{1}{c}{$\mu_{\delta}$} & \multicolumn{1}{c}{${{\mu}^\prime_{\alpha^{*}}}$} &  \multicolumn{1}{c}{${\mu}^\prime_{\delta}$}\\
		  & \multicolumn{1}{c}{(kpc)} & & & \multicolumn{1}{c}{(mas yr$^{-1}$)} & \multicolumn{1}{c}{(mas yr$^{-1}$)} & \multicolumn{1}{c}{(mas yr$^{-1}$)} & \multicolumn{1}{c}{(mas yr$^{-1}$)}\\
		\hline
		\hline

Reticulum II & 31.7 & 18 & 1143 & 2.479 $\pm$ 0.037 ($\pm$0.039)  & --1.268 $\pm$ 0.043 ($\pm$0.045)& 2.487 $\pm$ 0.037 ($\pm$0.040)  & --1.276 $\pm$ 0.043 ($\pm$0.046)\\ 
Bootes I & 64.0 & 53 & 1127 & --0.378 $\pm$ 0.024 ($\pm$0.027)  & --1.061 $\pm$ 0.018 ($\pm$0.023)& --0.357 $\pm$ 0.026 ($\pm$0.029)  & --1.071 $\pm$ 0.020 ($\pm$0.024)\\ 
Sagittarius II & 66.0 & 20 & 273 & --0.777 $\pm$ 0.061 ($\pm$0.063)  & --0.842 $\pm$ 0.038 ($\pm$0.041)& --0.788 $\pm$ 0.063 ($\pm$0.065)  & --0.841 $\pm$ 0.039 ($\pm$0.042)\\ 
Draco & 75.9 & 434 & 1261 & 0.035 $\pm$ 0.007 ($\pm$0.014)  & --0.180 $\pm$ 0.008 ($\pm$0.014)& 0.033 $\pm$ 0.011 ($\pm$0.016)  & --0.183 $\pm$ 0.011 ($\pm$0.017)\\ 
Ursa Minor & 77.9 & 728 & 1154 & --0.116 $\pm$ 0.006 ($\pm$0.009)  & 0.076 $\pm$ 0.006 ($\pm$0.010)& --0.121 $\pm$ 0.010 ($\pm$0.012)  & 0.080 $\pm$ 0.010 ($\pm$0.012)\\ 
Sculptor & 85.9 & 3119 & 1110 & 0.093 $\pm$ 0.003 ($\pm$0.010)  & --0.159 $\pm$ 0.002 ($\pm$0.009)& 0.081 $\pm$ 0.007 ($\pm$0.011)  & --0.156 $\pm$ 0.007 ($\pm$0.011)\\ 
Sextans & 89.0 & 244 & 526 & --0.387 $\pm$ 0.013 ($\pm$0.014)  & 0.038 $\pm$ 0.013 ($\pm$0.014)& --0.373 $\pm$ 0.018 ($\pm$0.019)  & 0.021 $\pm$ 0.019 ($\pm$0.020)\\ 
Ursa Major I & 101.6 & 11 & 1399 & --0.405 $\pm$ 0.056 ($\pm$0.058)  & --0.616 $\pm$ 0.065 ($\pm$0.067)& --0.406 $\pm$ 0.056 ($\pm$0.059)  & --0.612 $\pm$ 0.065 ($\pm$0.067)\\ 
Carina & 106.9 & 297 & 932 & 0.531 $\pm$ 0.009 ($\pm$0.014)  & 0.126 $\pm$ 0.009 ($\pm$0.014)& 0.532 $\pm$ 0.012 ($\pm$0.017)  & 0.127 $\pm$ 0.012 ($\pm$0.017)\\ 
Crater II & 116.4 & 35 & 650 & --0.113 $\pm$ 0.037 ($\pm$0.038)  & --0.086 $\pm$ 0.025 ($\pm$0.026)& --0.131 $\pm$ 0.039 ($\pm$0.041)  & --0.088 $\pm$ 0.026 ($\pm$0.027)\\ 
Fornax & 149.3 & 5488 & 1126 & 0.380 $\pm$ 0.001 ($\pm$0.005)  & --0.361 $\pm$ 0.002 ($\pm$0.006)& 0.386 $\pm$ 0.005 ($\pm$0.007)  & --0.367 $\pm$ 0.007 ($\pm$0.009)\\ 
C. Venatici I & 217.5 & 41 & 1311 & --0.137 $\pm$ 0.043 ($\pm$0.045)  & --0.108 $\pm$ 0.027 ($\pm$0.030)& --0.136 $\pm$ 0.043 ($\pm$0.046)  & --0.106 $\pm$ 0.028 ($\pm$0.031)\\ 
Leo II & 235.9 & 52 & 650 & --0.124 $\pm$ 0.049 ($\pm$0.051)  & --0.141 $\pm$ 0.047 ($\pm$0.049)& --0.127 $\pm$ 0.050 ($\pm$0.052)  & --0.135 $\pm$ 0.048 ($\pm$0.050)\\ 
Leo I & 257.4 & 294 & 594 & --0.060 $\pm$ 0.024 ($\pm$0.027)  & --0.112 $\pm$ 0.017 ($\pm$0.022)& --0.041 $\pm$ 0.027 ($\pm$0.030)  & --0.150 $\pm$ 0.020 ($\pm$0.024)\\ 

		\hline

	\end{tabular}
	\label{tab:resultadosPM}
\end{table*}
%\end{landscape}

In general, our results are consistent within $1\sigma$ with those from previous studies based on \textit{Gaia} astrometry (DR2: \citealt{Helmi2018, Fritz2018, MassariHelmi2018, Simon2018, PaceLi2019}; EDR3: \citealt{McConnachie2020b}), being most similar to those conducted using EDR3 data. Nevertheless, there are small differences in the final nominal values and their associated uncertainties between this and previous studies. Such differences can be explained by the fact that we have applied astrometric zero-point corrections to our data and propagated all known sources of error to the uncertainties of the final measurements, such as possible small-scale systematic errors.
If we do not apply such corrections and propagate the small-scale uncertainties, we obtain results that are fully compatible with those of \citet{McConnachie2020b}, with the only exception of Reticulum II, which shows a PM offset of $\sim 2.5\sigma$ in the RA direction. However, our results for this galaxy are based on only 18 member stars, and the lack of statistical stability or the presence in our sample of even just a few MW contaminants could be behind this discrepancy.

\begin{figure}
    \centering
    \includegraphics[width=\columnwidth,scale=0.58]{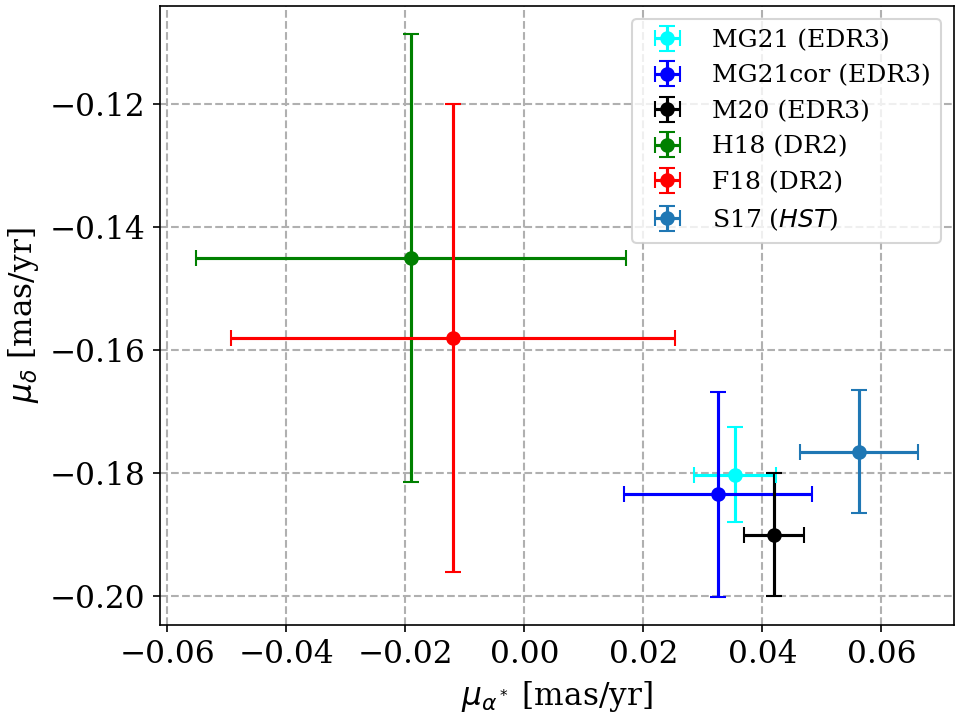}
    \caption{The PMs of Draco from this work compared to previous studies. MG21 (EDR3): this work considering random errors only; MG21cor (EDR3): this work including systematic errors and zero-point corrections; 
    M20 (EDR3): \citet{McConnachie2020b}; H18 (DR2): \citet{Helmi2018}; F18 (DR2): \citet{Fritz2018}; and
    S17 (\textit{HST}): \citet{Sohn2017}. We consider a systematic uncertainty of 0.035 mas yr$^{-1}$ for F18 and H18 (\citealt{Fritz2018, Helmi2018}).
    }
    \label{fig:draco}
\end{figure}

Our results are also broadly consistent with those derived using the {\it Hubble Space Telescope} (HST) in Draco and Leo I (\citealt{Sohn2013, Sohn2017}). However, in the case of Sculptor, we find a PM offset of $\sim 2.7\sigma$ in the RA direction. This offset greatly improves after applying the astrometric zero-point corrections to $\sim1.5 \sigma$, indicating that the EDR3 data may be strongly affected by non-zero astrometric zero-points in the region of Sculptor. In any case, possible large-scale variations in the astrometric zero-points are not expected to have a large impact on the derived internal kinematics, since these affect all the stars in the galaxy fairly equally. Fig.~\ref{fig:draco} shows our PMs derived for Draco compared to those from previous works available in the literature.

\subsection{Internal kinematics}

\subsubsection{General trends}
\label{subsec:genten}
We derived the mean radial ($V_R$, i.e. expansion/contraction) and tangential ($V_T$, i.e. rotation) components of the velocity in the plane of the sky following the procedure described in Section~\ref{sec:v1v2v3}. We considered deriving the velocity dispersion along these two components. However, this would require a somewhat sophisticated modelling of the observed scatter, given the large errors per star per component ($\gtrsim 100 \kms$ on average in all our galaxies) and our lack of knowledge about possible small-scale systematic errors. Instead, we decided to use the line-of-sight velocity dispersion values found in \citet{McConnachie2020a}. While the observed velocity dispersion is not necessarily the same along all viewing directions, we expect its values to be reasonably consistent, given the mostly pressure-supported nature of these systems. Furthermore, dispersion velocities measured on the plane of the sky seem to roughly coincide with those measured along the line-of-sight in Draco and Sculptor (\citealt{Massari2018, Massari2020}).

The results for our \textit{sample 2}, i.e. galaxies whose uncertainties in the mean velocity components are below the internal velocity dispersion, are listed in Table~\ref{tab:resultadosv}. Net positive or negative values of $V_R$ imply, respectively, the expansion or contraction of the galaxy in the plane of the sky. Net values of $V_T$ indicate rotation on the plane of the sky, with positive values corresponding to the anticlockwise direction.

\begin{table*}
    \centering
    \caption{Kinematic results for \textit{sample 2}, formed by the six galaxies for which we were able to study internal kinematics. Columns: (1) galaxy name, (2) projected ellipticity (\citealt{McConnachie2020a}), (3) derived projected ellipticity (see Section~\ref{subsec:genten}), (4) velocity dispersion (\citealt{McConnachie2020a}), (5) dynamic mass within the half-light radius (see Section~\ref{subsec:genten}), (6) mean radial velocity in the plane of the sky, (7) mean tangential velocity in the plane of the sky, and (8) ratio rotation velocity to velocity dispersion. $\dagger$, estimated in this work}
	\begin{tabular}{l|r|r|r|r|r|r|c} 
	    \hline
		Galaxy & \multicolumn{1}{c}{$e$} & \multicolumn{1}{c}{$e^{\dagger}$}&  \multicolumn{1}{c}{$\sigma_v$}  & \multicolumn{1}{c}{$M(r_h)^{\dagger}$} & \multicolumn{1}{c}{$V_R^{\dagger}$} & \multicolumn{1}{c}{$V_T^{\dagger}$} & \multicolumn{1}{c}{$ {\left| {V}_T\right| / \sigma_v}^{\dagger}$}\\
		& &  & \multicolumn{1}{c}{(km s$^{-1}$)} & \multicolumn{1}{c}{($\times 10^7 M_{\odot}$)} & \multicolumn{1}{c}{(km s$^{-1}$)} & \multicolumn{1}{c}{(km s$^{-1}$)} & \\
		\hline
		\hline
Draco      & 0.30 $\pm$ 0.01 & 0.27 &  $9.1 \pm 1.2$ & 1.05 $\pm$ 0.29 & 0.99 $\pm$ 2.71 & 0.95 $\pm$ 2.70 & 0.10 $\pm$ 0.30\\ 
Ursa Minor & 0.55 $\pm$ 0.01 & 0.48 & $9.5 \pm 1.2$ & 2.00 $\pm$ 0.51 & 1.36 $\pm$ 2.23 & 0.03 $\pm$ 2.28 & 0.003 $\pm$ 0.240\\
Sculptor   & 0.37 $\pm$ 0.01 & 0.17 & $9.2 \pm 1.4$ &  1.51 $\pm$ 0.47 & -0.22 $\pm$ 1.05 & -3.00 $\pm$ 1.02 & 0.33 $\pm$ 0.12\\ 
Sextans    & 0.35 $\pm$ 0.05 & 0.20&  $7.9 \pm 1.3$ & 2.51 $\pm$ 0.84 & -0.69 $\pm$ 5.35 & -0.61 $\pm$ 5.24 & 0.08 $\pm$ 0.66\\ 
Carina     & 0.37 $\pm$ 0.01 & 0.32 & $6.6 \pm 1.2$ & 0.88 $\pm$  0.33 & 4.66 $\pm$ 4.50 & -9.56 $\pm$ 4.50 & 1.45 $\pm$ 0.73\\ 
Fornax     & 0.28 $\pm$ 0.01 & 0.26 & $11.7 \pm 0.9$ & 6.26 $\pm$ 1.10 & -0.88 $\pm$ 1.32 & -2.77 $\pm$ 1.30 & 0.24 $\pm$ 0.11\\ 

		\hline
	\end{tabular}
	\label{tab:resultadosv}
\end{table*}

No net radial expansion or contraction was detected in our \textit{sample 2} galaxies, except perhaps in the case of Carina, which shows expansion at the $\sim 1\sigma$ level. In turn, $V_T$ values and their errors show evidence for clockwise rotation in Fornax, Sculptor and Carina at the 2.1, 2.9, and 2.1$\sigma$ levels, respectively.

\begin{figure}
    \centering
    \includegraphics[width=\columnwidth,scale=0.28]{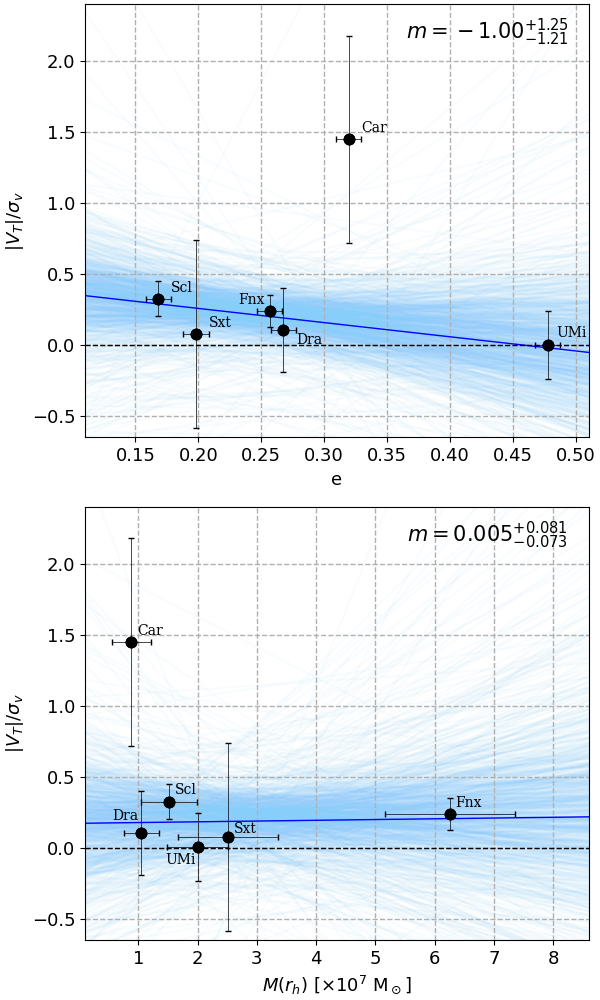}
    \caption{Rotation support ratio, $|V_T| / \sigma_v$, versus the ellipticity and dynamical mass. Top panel: $|V_T| / \sigma_v$ as a function of the projected ellipticity, $e$. Bottom panel: $|V_T| / \sigma_v$ as a function of the dynamical mass within the half-light radius, $M(r_h)$. The values of $\sigma_v$ are along the line of sight and are taken from \citet{McConnachie2020a} (see Section~\ref{subsec:genten}). Dark blue lines represent the best linear fits obtained with an MCMC. Light blue lines show a thousand random samples from the MCMC chain. Carina has been excluded from the fits.
    }
    \label{fig:internal_properties}
\end{figure}

The detected rotation signal is the projection in the celestial sphere of the internal dynamics of the galaxies and could be related to other internal properties. To investigate such possible relationships, we used line-of-sight velocity dispersion values, $\sigma_v$, from the literature (\citealt{McConnachie2020a}) to derive $|V_T|/\sigma_v$ for all the galaxies in our \textit{sample 2}. Here, $|V_T|/\sigma_v > 1$ indicates significant rotation support. We derived the projected ellipticity ($e$) through the Multi-Gaussian Expansion parametrization of the galaxies' surface brightness with the fitting algorithm of
\citet{Cappellari2002}. These ellipticities are generally similar to previously reported values (\citealt{McConnachie2020a}), being most different in the cases of Sculptor and Sextans, which could be due to contamination from MW stars in previous studies. We also derived the dynamical mass within the half-light radius using the relation $M(r_h) = 580r_h\sigma_v^2$ from  \citet{Walker2009sig}. These quantities can be found in Table~\ref{tab:resultadosv} and are shown and compared in Fig.~\ref{fig:internal_properties}, where linear fittings using Markov Chain Monte Carlo (MCMC) are shown by dark blue lines. 

The case of Carina deserves some special attention. With $|V_T| = 9.56\pm 4.50 \kms$ and $|V_T|/\sigma_v = 1.45 \pm 0.73$, the galaxy seems to be an outlier (at $\sim 2 \sigma$ level) in a sample of otherwise pressure-supported galaxies. Interestingly, no internal rotation from line-of-sight velocity gradients has been reported up to date in the core of Carina \citep{Wheeler2017}, and it is the galaxy with the lowest line-of-sight velocity dispersion in our \textit{sample 2}. If the values obtained here are accurate, Carina would be rotating in the plane of the sky with its internal angular momentum pointing along the line of sight and thus a large part of its 1D dispersion signal would also be contained in the same plane. This would imply a significantly larger mass for Carina. Assuming spherical symmetry, the addition of rotation to the mass calculation yields values around $\sim2\times10^8 M_\odot$.

On the other hand, Carina has, together with Sextans, the smallest number of identified members in our \textit{sample 2}. That, the large errors in $V_T$, and the fact that Carina shows the largest contamination in our \textit{sample 2} (6.3\%, see Section~\ref{sec:Samples_of_galaxies}) could cast some doubts on the reliability of the results obtained here. In addition, the PMs used by \citet{Wheeler2017} to correct for projection effects in their $ \vlos $ gradients were derived by \citet{Piatek2003}, which are not consistent in the Dec. direction with those derived here or others works based on Gaia data. The use of more recent PM values could introduce changes in the inferred velocity gradients, which in turn could result in the detection of rotation along the line of sight.

Either way, being an outlier or having unreliable measurements, Carina deserves further individual study that is beyond the scope of this work, and it seems sensible to not include it in our fittings. Therefore, we decided to exclude Carina from our MCMC experiments, but to continue showing its values throughout the paper. Lastly, it is worth noticing that even multiplying the error bars by two, Carina would still exhibit clockwise rotation, which leads us to confirm such rotation with the present data.

For the rest of the galaxies, we measure a slope of $ -1.00^{+1.25}_{-1.21}$ between $|V_T|/\sigma_v$ and ellipticity; this hints at a mild anticorrelation, but is also consistent with no trend within the uncertainties, so it is not a significant detection (at $\sim 0.8 \sigma$ level).
For the fit, we assumed the uncertainties of the ellipticity to be the median value of the uncertainties of ellipticities from the literature (0.01, \citealt{McConnachie2020a}). Assuming no errors for the ellipticity yields almost the same slope, $-1.01^{+1.25}_{-1.21}$. Repeating the fit using the ellipticities from the literature, we obtained a similar result, $-0.61^{+1.56}_{-1.63}$.

Here, a negative sign in the slope indicates that galaxies showing larger $|V_T|/\sigma_v$ tend to have lower projected ellipticities. The tendency, albeit weak, could be related to the flattening of the galaxies along their axes of rotation. Galaxies with larger projected rotation on the sky, i.e. oblate systems with larger values of $|V_T|/\sigma_v$, are expected to have their rotation axes closer to perpendicular to the plane of the sky, thus showing some flattening along the line of sight and smaller sky-projected ellipticities. The different ellipticity values observed would be, therefore, caused by the different random orientation of the galaxies around the MW. However, given the small number of galaxies in our sample, and the large uncertainties in the slope of the linear fit, we do not make any strong claim about this particular subject. Furthermore, systems deformed by their intrinsic rotation are expected to show much larger $|V_T|/\sigma_v$ values than the ones observed here \citep{Binney1978, Wheeler2017}. We therefore conclude that the rotation in the plane of the sky is not the main mechanism behind the projected shape of these galaxies.

We also searched for but did not detect any trend ($\sim 0.06 \sigma$) between $|V_T|/\sigma_v$ and the total mass enclosed within the half-light radius, $M(r_h)$. We repeated the experiment considering the total mass estimations from \citet{Lokas2009}, obtaining similar results. This was also found by \citet{Wheeler2017} and indicates that, in general, galaxies with higher rotation velocities have also higher dispersion velocities and larger masses.

In light of these results, we conclude that Fornax, Sculptor and Carina show rotation in the plane of the sky. We also confirm that random motions are the main contribution to the internal velocities and ellipsoid shape of the galaxies in our \textit{sample 2}.

\subsubsection{Velocity profiles and kinematic maps}

To study the spatially resolved internal kinematics, we further restrict our galaxy sample to those containing at least 500 member stars, to ensure statistically sound results. Therefore, in this section we study Fornax, Sculptor, and Ursa Minor (\textit{sample 3}). The quality of our data in these galaxies allows us to study their internal kinematics averaged at different galactocentric distances, position angles, and 2D positions in the sky. To compute the averaged velocity components, we use three different binning schemes in each galaxy: a 1D binning in radius, a 1D binning in angular position, and a 2D binning in radius and angular position. For the first one, we use bins defined by fifth quantiles in galactocentric elliptical radius; for the second one, four equal angular sections starting from the galaxy optical minor axis; and for the third one, Voronoi cells (\citealt{Capellari2003, delPino2017}) of $\sim$ 500 stars in Fornax and Sculptor and $\sim$250 stars in Ursa Minor. For each bin in each binning scheme, we compute error-weighted averages of the stellar velocity components together with their corresponding standard errors (Eq.~\ref{eq:weightedaverages}). We denote $\tilde v_{R,j}$ and $\tilde v_{T,j}$ the average values for bin $j$, within each binning scheme. The total error within each Voronoi cell ($\Delta \tilde v_j$) is computed as the square root of the quadratic sum of the error of both velocity components.

\begin{figure*}
    \centering
    \includegraphics[width=1.0\textwidth,scale=0.17]{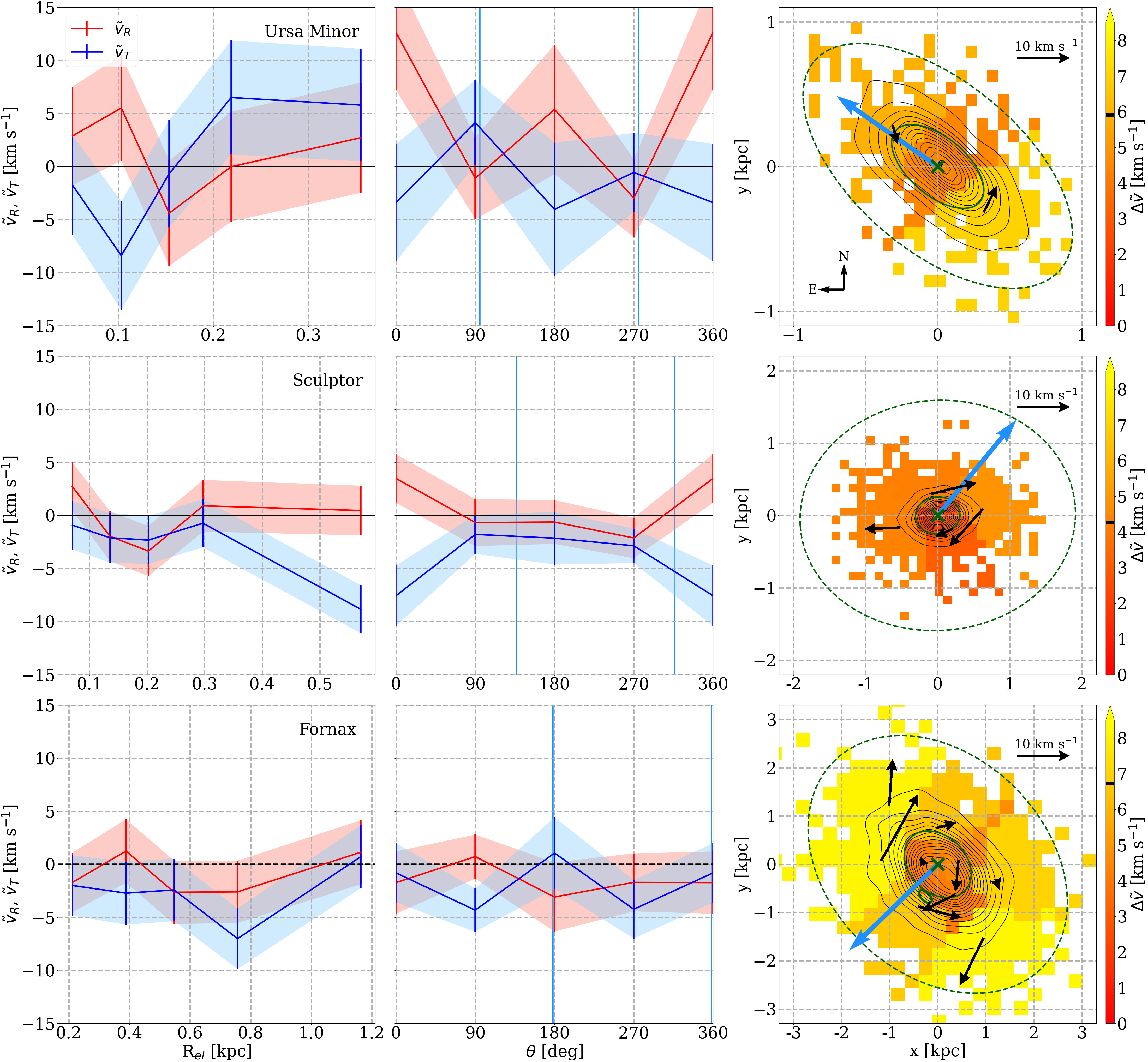}
    \caption{The spatially resolved internal kinematics of the Ursa Minor, Sculptor, and Fornax dSphs. From left to right, the three columns show the error-weighted mean of the tangential and radial components of the velocity $(\tilde v_{R}$, $\tilde v_{T})$ using the three different schemes described in the text. The first column shows $\tilde v_{R}$ in red, and $\tilde v_{T}$ in blue, computed within radial bins along the elliptical radii defined by the fifth quantile. The second column shows the same quantities, but derived within four angular sectors starting from the semiminor axis found in anticlockwise direction from the North direction, i.e. located in the East half of the galaxy. Errors are shown by the shaded areas in the corresponding colours. Blue vertical lines represent the direction of the PM in GSR in the second column. The third column shows the mean sky-projected 2D kinematics. Here, small black arrows show the mean internal velocities measured within the Voronoi cells of $\sim250$, $\sim500$, and $\sim500$ stars for Ursa Minor, Sculptor, and Fornax, respectively. The length of the arrows indicate the absolute value of the velocity, and the scale is the same for all panels. The median error of the mean internal velocities across cells is marked in the corresponding color bars with a black horizontal line. Blue arrows indicate the direction of the galaxies' systemic PM in the GSR. The projected stellar densities are shown by black contours. In the panel of Fornax, the position of the shell reported by \citet{Coleman2004} is represented by a small green ellipse located south-east from the centre of the galaxy. The rest of the markers coincide with those of Fig.~\ref{fig:member_selection}.
    }
    \label{fig:resolved_internal_kinematics}
\end{figure*}

In Fig.~\ref{fig:resolved_internal_kinematics}, we show (from left to right) the mean velocity components and their corresponding errors within the radial bins, the angular bins, and the Voronoi cells. In the rightmost panels, the systemic PMs are represented by large blue arrows and are referred to the Galactic Standard of Rest (GSR), i.e. after subtracting the reflex motion of the Sun around the MW. The averages of both internal velocity components, shown by black arrows, are referred to the CM of each galaxy. In the following, we analyse the main features for each galaxy.

Ursa Minor has the largest error bars both in the radial and angular profiles ($\sim$ 5 km s$^{-1}$). Nevertheless, the errors are below the galaxy velocity dispersion, $\sigma_v$, in all the bins. Radial and angular profiles show mean velocities mostly compatible with zero at $1 \sigma$, with an increase in $\tilde v_R$ along the semiminor axis (Fig.~\ref{fig:resolved_internal_kinematics}, second panel). The galaxy shows a weak anticlockwise rotation (Fig. ~\ref{fig:resolved_internal_kinematics}, third panel), although the velocity components are below their uncertainties ($\sim0.3 \sigma$) in the north-east cell, being consistent with $V_T \sim 0$. 

Sculptor shows a consistent clockwise rotation signal, $\tilde v_T < 0$, within all sampled bins and no net radial expansion or contraction, $\tilde v_R \simeq 0$. This rotation pattern can be more easily observed in rightmost panel of Fig.~\ref{fig:resolved_internal_kinematics}, and is compatible with Sculptor's mean tangential velocity, $V_T = -3.00\pm1.02 \kms$.

Overall, Fornax displays clockwise rotation, $\tilde v_T < 0$, and no net radial expansion or contraction, $\tilde v_R \simeq 0$, at all radii. Both components of the velocity show some dependence with the PA, with positive values of $\tilde v_T$, and negative values of $\tilde v_R$ along the south-east semiminor axis, $\theta \sim 180 \deg$. This can also be observed in the Voronoi tessellation maps in the rightmost panel of Fig.~\ref{fig:resolved_internal_kinematics}, with the two cells located south-east from the centre of Fornax showing apparent contraction (at $\sim 0.3 \sigma$ and $1.3 \sigma$, respectively) and one showing strong counter rotation (at $\sim2 \sigma$).

The case of Fornax is especially interesting for a number of reasons. The projected major axis and the systemic PM point in perpendicular directions. This strengthens previous detections of velocity gradients along its projected major axis (\citealt{Battaglia2006, delPino2017}), pointing to the presence of internal rotation along the line of sight. Apart from harbouring five globular clusters, Fornax is known to have substructures and shell-like clumps of stars in its young populations (\citealt{Coleman2004, Coleman2005, deBoer2013}) and to show asymmetries in the spatial distribution of its stellar populations (\citealt{Battaglia2006, DelPino2015, Wang2019}). We detect strong counterrotation ($\sim 2\sigma$) in a cell south-east from the galaxy centre, indicating that some of its stellar content may not be virialized. 
Interestingly, this cell contains the stellar shell found by \citet{Coleman2004}. 
These findings could be explained only by processes capable of asymmetrically perturbing the system, such as mergers with other systems (\citealt{Amorisco2012, DelPino2015, delPino2017}), or the accretion of previously expelled gas due to secular star formation processes.

\subsubsection{Environmental effects}
\label{sec:env}

Interaction with the MW is expected to have had an impact on the evolution of the satellites. In the context of the tidal stirring scenario (\citealt{Mayer2010} and references therein), rotation-supported dwarfs are tidally shocked in successive pericentre passages to be gradually transformed into pressure-supported dSphs, causing a drop in the internal rotation of dwarfs closer to the MW. Therefore, if this scenario is correct, the mostly pressure-supported dSphs we observe today would be the evolved state of former rotation-supported systems that have retained a fraction of their original rotation.

We have investigated possible traces of such interaction in the internal kinematics of the galaxies in our sample. To do so, we have studied their mean internal velocity components and their positions with respect to the MW. We have derived their Galactocentric distances assuming the Galactic centre to be located at $\alpha = 266.4051$ degrees, $\delta = $--28.936175 degrees (\citealt{Reid2004}), with a distance of 8.3 kpc (\citealt{Gillessen2009}), and a height of the Sun above the Galactic midplane of 27 pc (\citealt{Chen2001}).

\begin{figure}
    \centering
    \includegraphics[width=\columnwidth, scale=0.4]{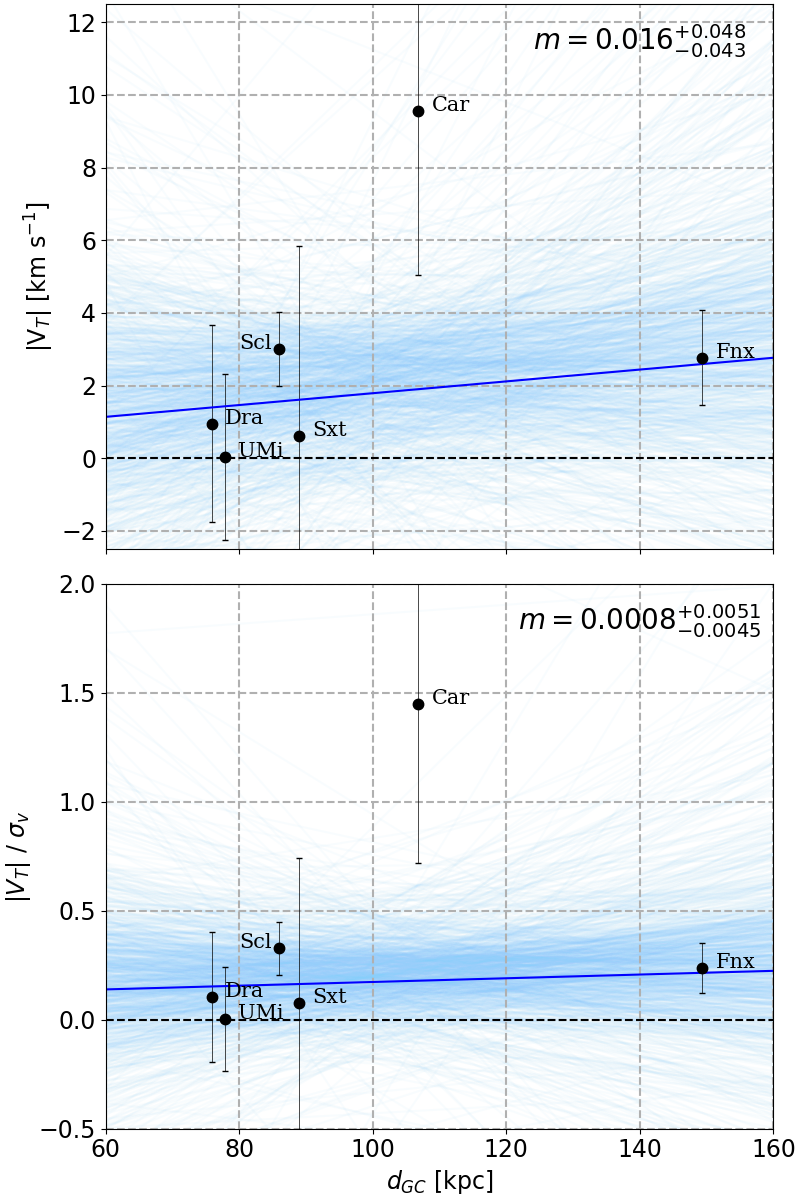}
    \caption{Internal kinematics of the galaxies as a function of their Galactocentric distance. Top panel: absolute values of the error-weighted mean of the rotation velocities, $\left|V_T\right|$, as a function of the Galactocentric distance, $d_{GC}$. Bottom panel: rotation support ratio,  $\left| {V}_T\right| / \sigma_v$,  as a function of $d_{GC}$. The values of $\sigma_v$ are along the line of sight and are taken from \citet{McConnachie2020a} (see Section~\ref{subsec:genten}). Dark blue lines represent the best linear fits obtained with an MCMC. Light blue lines show a thousand random samples from the MCMC chain. Carina has been excluded from the fits.
    }
    \label{fig:rotation_support_galactocentric_distance}
\end{figure}

Fig.~\ref{fig:rotation_support_galactocentric_distance} shows the sky-projected internal rotation for the galaxies in the \textit{sample 2} versus their Galactocentric distances. The top panel shows the absolute value of the rotation, $\left| {V}_T\right|$. Galaxies with lower $\left| {V}_T\right|$ tend to be located at lower Galactocentric distances. However, MCMC fitting yields a slope $m=0.02^{+0.05}_{-0.04} \kms {\mathrm{kpc}}^{-1}$, which is compatible with zero at 1 sigma (note that Carina is not included in the MCMC fit, as discussed in Section~\ref{subsec:genten}). Furthermore, the rotation support ratio, $\left| {V}_T\right| / \sigma_v$, also shows no significant trend with respect to the distance to the MW ($<0.2 \sigma$). This can be seen in the bottom panel of Fig.~\ref{fig:rotation_support_galactocentric_distance}, and indicates that, while farther galaxies could show slightly larger rotation signal, the overall rotation support remains mostly constant at all Galactocentric distances. In principle, this differs from tidal stirring models predictions where an increase of $\left| {V}_T\right| / \sigma_v$ is expected with increasing distance from the MW.

Nevertheless, it is important to point out that all the galaxies studied in this work lie within the virial radius of the MW, and that all of them are expected to have suffered several pericentres passages in a Hubble time. This could have significantly reduced the $\left| {V}_T\right| / \sigma_v$ ratio in all of them up to the point to make them indistinguishable one from another in terms of their rotation support. Also, the comparisons provided here use the galaxies' current Galactocentric distance as a proxy for the strength of the tidal interactions and thus for the efficiency of the possible dIrr to dSph transformation. In reality, this would depend on the number of passages and the pericentre distance, as well as a number of internal and external parameters of the dwarfs that have not been taken into account here e.g. the concentration of the dwarfs or the orientation of their internal angular momentum with respect to their orbital plane. Lastly, it is worth noticing that a straight line may not be the correct model to fit to these data.

Given all these caveats, we believe that no strong claim about the transformation of dwarfs from dIrr to dSph due to interactions with the MW can be drawn from these data. If anything, our results could suggest that tidal stirring was either a weak mechanism involved in the evolution of these particular systems or that it affected all the galaxies in our sample in a similar way (except Carina). These results are consistent with those found by \citet{Wheeler2017} who, analysing line-of-sight velocity gradients in a much larger sample of galaxies, did not find any strong trend between the rotation support and the distances of the dwarfs to their massive hosts. However, we stress the small number of galaxies used in this study and the fact that only the sky-projected components of the velocity are being considered. Future studies including more galaxies, velocity dispersion in the plane of the sky, or the three components of the stellar velocities may shed some light on this matter.

\section{Conclusions}
\label{sec:concl}
In this paper, we have explored the kinematics of 58 satellites of the MW using Gaia EDR3. We selected member stars using \textsc{GetGaia}, and derived systemic PMs for 14 of them, namely those containing more than 10 identified members. We accounted for the effects of possible small-scale systematic errors and local astrometric zero-points. Our results are in agreement with previous studies based on \textit{Gaia} and \textit{HST}, being generally compatible at 1$\sigma$ level. 

Combining the information from the PMs with line-of-sight velocities  
available in the literature, we derived the systemic 3D motion of these galaxies. Six of them, namely Carina, Draco, Fornax, Sculptor, Sextans, and Ursa Minor, have mean velocity component errors smaller than the corresponding line-of-sight velocity dispersion. The number of identified member stars ranged from 244 (Sextans) to 5488 (Fornax). We studied the internal kinematics of these six galaxies, projecting the stellar PMs into radial (expansion/contraction), $V_R$, and tangential (rotation), $V_T$, components with respect to their CM. 

We found no evidence for expansion or contraction ($V_R \neq 0 \kms$) in the studied galaxies, except perhaps for Carina, which shows expansion at $1 \sigma$ level.
The mean tangential component of the velocity in the plane of the sky, $V_T$, reveals significant rotation for Carina, Fornax, and Sculptor in the clockwise direction. Apart from that in Sagittarius dSph \citep{delPino2021}, these are the first measurements of internal rotation in the plane of the sky for dSphs satellites of the MW. All the galaxies show rotation velocities smaller than their velocity dispersion, $\left| {V}_T\right| / \sigma_v < 1$, except Carina albeit affected by a large error. 

Comparing the mean internal rotation of the dwarfs to their internal properties, we found that all the galaxies in our sample show values of $\left| {V}_T\right| / \sigma_v$ below that expected for systems dominated by their intrinsic rotation. While galaxies with lower $\left| {V}_T\right| / \sigma_v$ tend to show larger projected ellipticity, we did not find this trend to be significant, indicating that the rotation in the plane of the sky does not play an important role in the projected shape of these galaxies. We did not find any trends between $\left| {V}_T\right| / \sigma_v$ and the derived masses within the half-light radius or the total mass (\citealt{Lokas2009}) of these systems. On average, galaxies with larger rotation velocity also show larger velocity dispersion and masses.

We studied the spatially resolved internal kinematics of galaxies with at least 500 member stars, which constrained the sample to three dSphs: Fornax, Sculptor, and Ursa Minor. We observed clear rotation in Sculptor and Fornax. In Fornax, we found hints of expansion along the optical major axis at large radii and counterrotation in the region where the stellar shell found by \citet{Coleman2004} lies. 
This may indicate that some of the stellar content in Fornax is not virialized, which would be consistent with previous works suggesting that Fornax may have suffered a merger (\citealt{Amorisco2012, DelPino2015, delPino2017}), or the accretion of previously expelled local gas.

Finally, we studied how possible interactions of the satellites with the MW may have affected their internal kinematics. We found that galaxies with lower $|V_T|$ tend to be located at smaller Galactocentric distances. However, MCMC fittings of a linear model show that the trend is not significant. This is also the case for $\left| {V}_T\right| / \sigma_v$ and the Galactocentric distance, which remains roughly constant. The results presented here point to either very weak tidal stirring effects, or to a consistent tidal transformation for all the studied galaxies. Nevertheless, it is worth noticing that only the sky-projected components of the velocity of five dwarfs were used to draw these conclusions. Future studies analysing the full 3D internal kinematics of more galaxies may shed light on the dynamical state of these galaxies and on the mechanisms involved in their evolution.

\section*{Data availability}
All the data underlying this article are publicly available. The \textit{Gaia} EDR3 data can  be extracted from the \textit{Gaia} archive (\url{https://gea.esac.esa.int/archive}). The software for the membership selection is available at GitHub (\url{https://github.com/AndresdPM/GetGaia}). 
Catalogues including radial and tangential velocities for member stars will be shared on reasonable request to the corresponding author.

\section*{Acknowledgements}

We acknowledge support from the Spanish Ministry of Economy and Competitiveness (MINECO) under grant AYA2017-89841-P. The authors acknowledge grant support for program AR-15633 from the Space Telescope Science Institute (STScI), which is operated by the Association of Universities for Research in Astronomy (AURA), Inc., under NASA contract NAS5-26555. This project is part of the HSTPROMO (High-resolution Space Telescope PROper MOtion) Collaboration\footnote{https://www.stsci.edu/$\sim$marel/hstpromo.html}, a set of projects aimed at improving our dynamical understanding of stars, clusters and galaxies in the nearby Universe through measurement and interpretation of PMs from HST, \textit{Gaia}, and other space observatories. We thank the collaboration members for the sharing of their ideas and software. The authors also acknowledge all the open-source software involved in this study, specially TOPCAT, Python, PostgreSQL and git. This work has made use of data from the European Space Agency (ESA) mission {\it Gaia} (\url{https://www.cosmos.esa.int/gaia}), processed by the {\it Gaia} Data Processing and Analysis Consortium (DPAC, \url{https://www.cosmos.esa.int/web/gaia/dpac/consortium}). Funding for the DPAC has been provided by national institutions, in particular, the institutions participating in the {\it Gaia} Multilateral Agreement.

%%%%%%%%%%%%%%%%%%%%%%%%%%%%%%%%%%%%%%%%%%%%%%%%%%

%%%%%%%%%%%%%%%%%%%% REFERENCES %%%%%%%%%%%%%%%%%%

% The best way to enter references is to use BibTeX:

\bibliographystyle{mnras}
\bibliography{biblio}

\begin{thebibliography}{}
\makeatletter
\relax
\def\mn@urlcharsother{\let\do\@makeother \do\$\do\&\do\#\do\^\do\_\do\%\do\~}
\def\mn@doi{\begingroup\mn@urlcharsother \@ifnextchar [ {\mn@doi@}
  {\mn@doi@[]}}
\def\mn@doi@[#1]#2{\def\@tempa{#1}\ifx\@tempa\@empty \href
  {http://dx.doi.org/#2} {doi:#2}\else \href {http://dx.doi.org/#2} {#1}\fi
  \endgroup}
\def\mn@eprint#1#2{\mn@eprint@#1:#2::\@nil}
\def\mn@eprint@arXiv#1{\href {http://arxiv.org/abs/#1} {{\tt arXiv:#1}}}
\def\mn@eprint@dblp#1{\href {http://dblp.uni-trier.de/rec/bibtex/#1.xml}
  {dblp:#1}}
\def\mn@eprint@#1:#2:#3:#4\@nil{\def\@tempa {#1}\def\@tempb {#2}\def\@tempc
  {#3}\ifx \@tempc \@empty \let \@tempc \@tempb \let \@tempb \@tempa \fi \ifx
  \@tempb \@empty \def\@tempb {arXiv}\fi \@ifundefined
  {mn@eprint@\@tempb}{\@tempb:\@tempc}{\expandafter \expandafter \csname
  mn@eprint@\@tempb\endcsname \expandafter{\@tempc}}}

\bibitem[\protect\citeauthoryear{Amorisco \& Evans}{Amorisco \&
  Evans}{2012}]{Amorisco2012}
Amorisco N.~C.,  Evans N.~W.,  2012, \mn@doi [\mnras]
  {10.1111/j.1365-2966.2011.19684.x}, 419, 184

\bibitem[\protect\citeauthoryear{Aparicio, Carrera  \&
  Mart{\'{i}}nez-Delgado}{Aparicio et~al.}{2001}]{Aparicio2001}
Aparicio A.,  Carrera R.,   Mart{\'{i}}nez-Delgado D.,  2001, \mn@doi [\aj]
  {10.1086/323535}, 122, 2524

\bibitem[\protect\citeauthoryear{Battaglia et~al.,}{Battaglia
  et~al.}{2006}]{Battaglia2006}
Battaglia G.,  et~al., 2006, \mn@doi [\aap] {10.1051/0004-6361:20065720}, 459,
  423

\bibitem[\protect\citeauthoryear{Battaglia, Helmi, Tolstoy, Irwin, Hill  \&
  Jablonka}{Battaglia et~al.}{2008}]{Battaglia2008b}
Battaglia G.,  Helmi A.,  Tolstoy E.,  Irwin M.,  Hill V.,   Jablonka P.,
  2008, \mn@doi [\apj] {10.1086/590179}, 681, L13

\bibitem[\protect\citeauthoryear{Battaglia, Tolstoy, Helmi, Irwin, Parisi, Hill
   \& Jablonka}{Battaglia et~al.}{2011}]{Battaglia2011}
Battaglia G.,  Tolstoy E.,  Helmi A.,  Irwin M.,  Parisi P.,  Hill V.,
  Jablonka P.,  2011, \mn@doi [\mnras] {10.1111/j.1365-2966.2010.17745.x}, 411,
  1013

\bibitem[\protect\citeauthoryear{Belokurov et~al.,}{Belokurov
  et~al.}{2007}]{Belokurov2007}
Belokurov V.,  et~al., 2007, \mn@doi [\apj] {10.1086/509718}, 654, 897

\bibitem[\protect\citeauthoryear{Bettinelli, Hidalgo, Cassisi, Aparicio  \&
  Piotto}{Bettinelli et~al.}{2018}]{Bettinelli2018}
Bettinelli M.,  Hidalgo S.~L.,  Cassisi S.,  Aparicio A.,   Piotto G.,  2018,
  \mn@doi [\mnras] {10.1093/mnras/sty226}, 476, 71

\bibitem[\protect\citeauthoryear{Bettinelli, Hidalgo, Cassisi, Aparicio,
  Piotto, Valdes  \& Walker}{Bettinelli et~al.}{2019}]{Bettinelli2019}
Bettinelli M.,  Hidalgo S.~L.,  Cassisi S.,  Aparicio A.,  Piotto G.,  Valdes
  F.,   Walker A.~R.,  2019, \mn@doi [\mnras] {10.1093/mnras/stz1679}, 487,
  5862

\bibitem[\protect\citeauthoryear{Binney}{Binney}{1978}]{Binney1978}
Binney J.,  1978, \mn@doi [\mnras] {10.1093/mnras/183.3.501}, 183, 501

\bibitem[\protect\citeauthoryear{Blumenthal, Faber, Primack  \&
  Rees}{Blumenthal et~al.}{1984}]{Blumenthal1984}
Blumenthal G.~R.,  Faber S.~M.,  Primack J.~R.,   Rees M.~J.,  1984, \mn@doi
  [Nature] {10.1038/311517a0}, 311, 517

\bibitem[\protect\citeauthoryear{Cappellari}{Cappellari}{2002}]{Cappellari2002}
Cappellari M.,  2002, \mn@doi [\mnras] {10.1046/j.1365-8711.2002.05412.x}, 333,
  400

\bibitem[\protect\citeauthoryear{Cappellari \& Copin}{Cappellari \&
  Copin}{2003}]{Capellari2003}
Cappellari M.,  Copin Y.,  2003, \mn@doi [\mnras]
  {10.1046/j.1365-8711.2003.06541.x}, 342, 345

\bibitem[\protect\citeauthoryear{Carrera, Aparicio, Mart{\'{i}}nez-Delgado  \&
  Alonso-Garc{\'{i}}a}{Carrera et~al.}{2002}]{Carrera2002}
Carrera R.,  Aparicio A.,  Mart{\'{i}}nez-Delgado D.,   Alonso-Garc{\'{i}}a J.,
   2002, \mn@doi [\aj] {10.1086/340702}, 123, 3199

\bibitem[\protect\citeauthoryear{Chen et~al.,}{Chen et~al.}{2001}]{Chen2001}
Chen B.,  et~al., 2001, \mn@doi [\apj] {10.1086/320647}, 553, 184

\bibitem[\protect\citeauthoryear{Coleman \& {Da Costa}}{Coleman \& {Da
  Costa}}{2005}]{Coleman2005}
Coleman M.~G.,  {Da Costa} G.~S.,  2005, \mn@doi [\pasa] {DOI:
  10.1071/AS04067}, 22, 162

\bibitem[\protect\citeauthoryear{Coleman, Costa, Bland-Hawthorn,
  Martnez-Delgado, Freeman  \& Malin}{Coleman et~al.}{2004}]{Coleman2004}
Coleman M.,  Costa G. S.~D.,  Bland-Hawthorn J.,  Martnez-Delgado D.,  Freeman
  K.~C.,   Malin D.,  2004, \mn@doi [\aj] {10.1086/381298}, 127, 832

\bibitem[\protect\citeauthoryear{Dekel \& Silk}{Dekel \&
  Silk}{1986}]{Dekel1986}
Dekel A.,  Silk J.,  1986, \mn@doi [ApJ] {10.1086/164050}, 303, 39

\bibitem[\protect\citeauthoryear{Fabricius et~al.,}{Fabricius
  et~al.}{2021}]{Fabricius2020}
Fabricius C.,  et~al., 2021, \mn@doi [A\&A] {10.1051/0004-6361/202039834}, 649,
  A5

\bibitem[\protect\citeauthoryear{Fritz, Battaglia, Pawlowski, Kallivayalil,
  van~der Marel, Sohn, Brook  \& Besla}{Fritz et~al.}{2018}]{Fritz2018}
Fritz T.~K.,  Battaglia G.,  Pawlowski M.~S.,  Kallivayalil N.,  van~der Marel
  R.,  Sohn S.~T.,  Brook C.,   Besla G.,  2018, \mn@doi [A{\&}A]
  {10.1051/0004-6361/201833343}, 619, A103

\bibitem[\protect\citeauthoryear{{Gaia Collaboration} et~al.,}{{Gaia
  Collaboration} et~al.}{2016}]{GaiaCollaboration2016}
{Gaia Collaboration} et~al., 2016, \mn@doi [\aap]
  {10.1051/0004-6361/201629272}, 595, A1

\bibitem[\protect\citeauthoryear{{Gaia Collaboration} et~al.,}{{Gaia
  Collaboration} et~al.}{2018}]{Helmi2018}
{Gaia Collaboration} et~al., 2018, \mn@doi [\aap]
  {10.1051/0004-6361/201832698}, 616, A12

\bibitem[\protect\citeauthoryear{{Gaia Collaboration} et~al.,}{{Gaia
  Collaboration} et~al.}{2021}]{GaiaEDR3}
{Gaia Collaboration} et~al., 2021, \mn@doi [A\&A]
  {10.1051/0004-6361/202039657}, 649, A1

\bibitem[\protect\citeauthoryear{{Gallagher} \& Wyse}{{Gallagher} \&
  Wyse}{1994}]{Gallagher1994}
{Gallagher} J. S. I. I.~I.,  Wyse R. F.~G.,  1994, \pasp, 106, 1225

\bibitem[\protect\citeauthoryear{Gillessen, Eisenhauer, Trippe, Alexander,
  Genzel, Martins  \& Ott}{Gillessen et~al.}{2009}]{Gillessen2009}
Gillessen S.,  Eisenhauer F.,  Trippe S.,  Alexander T.,  Genzel R.,  Martins
  F.,   Ott T.,  2009, \mn@doi [\apj] {10.1088/0004-637x/692/2/1075}, 692, 1075

\bibitem[\protect\citeauthoryear{Hidalgo et~al.,}{Hidalgo
  et~al.}{2013}]{Hidalgo2013}
Hidalgo S.~L.,  et~al., 2013, \mn@doi [\apj] {10.1088/0004-637x/778/2/103},
  778, 103

\bibitem[\protect\citeauthoryear{Irwin \& Hatzidimitriou}{Irwin \&
  Hatzidimitriou}{1995}]{Irwin1995}
Irwin M.,  Hatzidimitriou D.,  1995, \mn@doi [\mnras]
  {10.1093/mnras/277.4.1354}, 277, 1354

\bibitem[\protect\citeauthoryear{Kazantzidis, {\L}okas, Callegari, Mayer  \&
  Moustakas}{Kazantzidis et~al.}{2011}]{Kazantzidis2011}
Kazantzidis S.,  {\L}okas E.~L.,  Callegari S.,  Mayer L.,   Moustakas L.~A.,
  2011, \mn@doi [\apj] {10.1088/0004-637x/726/2/98}, 726, 98

\bibitem[\protect\citeauthoryear{Kirby, Bullock, Boylan-Kolchin, Kaplinghat  \&
  Cohen}{Kirby et~al.}{2014}]{Kirby2014}
Kirby E.~N.,  Bullock J.~S.,  Boylan-Kolchin M.,  Kaplinghat M.,   Cohen J.~G.,
   2014, \mn@doi [\mnras] {10.1093/mnras/stu025}, 439, 1015

\bibitem[\protect\citeauthoryear{Kleyna, Wilkinson, Evans  \& Gilmore}{Kleyna
  et~al.}{2004}]{Kleyna2004}
Kleyna J.~T.,  Wilkinson M.~I.,  Evans N.~W.,   Gilmore G.,  2004, \mn@doi
  [\mnras] {10.1111/j.1365-2966.2004.08434.x}, 354, 66

\bibitem[\protect\citeauthoryear{Kleyna, Wilkinson, Evans  \& Gilmore}{Kleyna
  et~al.}{2005}]{Kleyna2005}
Kleyna J.~T.,  Wilkinson M.~I.,  Evans N.~W.,   Gilmore G.,  2005, \apjl, 630,
  L141

\bibitem[\protect\citeauthoryear{Koch, Kleyna, Wilkinson, Grebel, Gilmore,
  Evans, Wyse  \& Harbeck}{Koch et~al.}{2007a}]{Koch2007b}
Koch A.,  Kleyna J.~T.,  Wilkinson M.~I.,  Grebel E.~K.,  Gilmore G.~F.,  Evans
  N.~W.,  Wyse R. F.~G.,   Harbeck D.~R.,  2007a, \mn@doi [\aj]
  {10.1086/519380}, 134, 566

\bibitem[\protect\citeauthoryear{Koch, Wilkinson, Kleyna, Gilmore, Grebel,
  Mackey, Evans  \& Wyse}{Koch et~al.}{2007b}]{Koch2007a}
Koch A.,  Wilkinson M.~I.,  Kleyna J.~T.,  Gilmore G.~F.,  Grebel E.~K.,
  Mackey A.~D.,  Evans N.~W.,   Wyse R. F.~G.,  2007b, \mn@doi [\apj]
  {10.1086/510879}, 657, 241

\bibitem[\protect\citeauthoryear{Koch, Grebel, Gilmore, Wyse, Kleyna, Harbeck,
  Wilkinson  \& Evans}{Koch et~al.}{2008}]{Koch2008}
Koch A.,  Grebel E.~K.,  Gilmore G.~F.,  Wyse R.~F.,  Kleyna J.~T.,  Harbeck
  D.~R.,  Wilkinson M.~I.,   Evans N.~W.,  2008, \mn@doi [\aj]
  {10.1088/0004-6256/135/4/1580}, 135, 1580

\bibitem[\protect\citeauthoryear{{Lindegren} et~al.,}{{Lindegren}
  et~al.}{2021a}]{Lindegren2020}
{Lindegren} L.,  et~al., 2021a, \mn@doi [\aap] {10.1051/0004-6361/202039709},
  \href {https://ui.adsabs.harvard.edu/abs/2021A&A...649A...2L} {649, A2}

\bibitem[\protect\citeauthoryear{{Lindegren} et~al.,}{{Lindegren}
  et~al.}{2021b}]{Lindegren2020b}
{Lindegren} L.,  et~al., 2021b, \mn@doi [\aap] {10.1051/0004-6361/202039653},
  \href {https://ui.adsabs.harvard.edu/abs/2021A&A...649A...4L} {649, A4}

\bibitem[\protect\citeauthoryear{{\L}okas}{{\L}okas}{2009}]{Lokas2009}
{\L}okas E.~L.,  2009, \mn@doi [Monthly Notices of the Royal Astronomical
  Society: Letters] {10.1111/j.1745-3933.2009.00620.x}, 394, L102

\bibitem[\protect\citeauthoryear{{\L}okas, Semczuk, Gajda  \&
  D'Onghia}{{\L}okas et~al.}{2015}]{Lokas2015}
{\L}okas E.~L.,  Semczuk M.,  Gajda G.,   D'Onghia E.,  2015, \mn@doi [\apj]
  {10.1088/0004-637x/810/2/100}, 810, 100

\bibitem[\protect\citeauthoryear{{Massari} \& {Helmi}}{{Massari} \&
  {Helmi}}{2018}]{MassariHelmi2018}
{Massari} D.,  {Helmi} A.,  2018, \mn@doi [\aap] {10.1051/0004-6361/201833367},
  620, A155

\bibitem[\protect\citeauthoryear{Massari, Breddels, Helmi, Posti, Brown  \&
  Tolstoy}{Massari et~al.}{2018}]{Massari2018}
Massari D.,  Breddels M.,  Helmi A.,  Posti L.,  Brown A.,   Tolstoy E.,  2018,
  Nature Astronomy, 2, 156

\bibitem[\protect\citeauthoryear{Massari, Helmi, Mucciarelli, Sales, Spina  \&
  Tolstoy}{Massari et~al.}{2020}]{Massari2020}
Massari D.,  Helmi A.,  Mucciarelli A.,  Sales L.~V.,  Spina L.,   Tolstoy E.,
  2020, \mn@doi [A\&A] {10.1051/0004-6361/201935613}, 633, A36

\bibitem[\protect\citeauthoryear{Mateo}{Mateo}{1998}]{Mateo1998}
Mateo M.,  1998, \mn@doi [\araa] {10.1146/annurev.astro.36.1.435}, 36, 435

\bibitem[\protect\citeauthoryear{Mayer}{Mayer}{2010}]{Mayer2010}
Mayer L.,  2010, \mn@doi [Advances in Astronomy] {10.1155/2010/278434}, 2010,
  278434

\bibitem[\protect\citeauthoryear{McConnachie}{McConnachie}{2012}]{McConnachie2012}
McConnachie A.~W.,  2012, \mn@doi [\aj] {10.1088/0004-6256/144/1/4}, 144

\bibitem[\protect\citeauthoryear{McConnachie \& Venn}{McConnachie \&
  Venn}{2020a}]{McConnachie2020b}
McConnachie A.~W.,  Venn K.~A.,  2020a, Research Notes of the AAS, 4, 229

\bibitem[\protect\citeauthoryear{McConnachie \& Venn}{McConnachie \&
  Venn}{2020b}]{McConnachie2020a}
McConnachie A.~W.,  Venn K.~A.,  2020b, \aj, 160, 124

\bibitem[\protect\citeauthoryear{Monelli et~al.,}{Monelli
  et~al.}{2010a}]{Monelli2010a}
Monelli M.,  et~al., 2010a, \mn@doi [\apj] {10.1088/0004-637x/720/2/1225}, 720,
  1225

\bibitem[\protect\citeauthoryear{Monelli et~al.,}{Monelli
  et~al.}{2010b}]{Monelli2010b}
Monelli M.,  et~al., 2010b, \mn@doi [\apj] {10.1088/0004-637x/722/2/1864}, 722,
  1864

\bibitem[\protect\citeauthoryear{Mu{\~{n}}oz et~al.,}{Mu{\~{n}}oz
  et~al.}{2005}]{Munoz2005}
Mu{\~{n}}oz R.~R.,  et~al., 2005, \mn@doi [\apj] {10.1086/497396}, 631, L137

\bibitem[\protect\citeauthoryear{Navarro, Frenk  \& White}{Navarro
  et~al.}{1995}]{NavarroFrenkWhite1995}
Navarro J.~F.,  Frenk C.~S.,   White S. D.~M.,  1995, \mn@doi [\mnras]
  {10.1093/mnras/275.3.720}, 275, 720

\bibitem[\protect\citeauthoryear{Okamoto, Arimoto, Yamada  \& Onodera}{Okamoto
  et~al.}{2012}]{Okamoto2012}
Okamoto S.,  Arimoto N.,  Yamada Y.,   Onodera M.,  2012, \mn@doi [\apj]
  {10.1088/0004-637x/744/2/96}, 744, 96

\bibitem[\protect\citeauthoryear{Pace \& Li}{Pace \& Li}{2019}]{PaceLi2019}
Pace A.~B.,  Li T.~S.,  2019, \mn@doi [\apj] {10.3847/1538-4357/ab0aee}, 875,
  77

\bibitem[\protect\citeauthoryear{Pace et~al.,}{Pace et~al.}{2020}]{Pace2020}
Pace A.~B.,  et~al., 2020, \mn@doi [\mnras] {10.1093/mnras/staa1419}, 495, 3022

\bibitem[\protect\citeauthoryear{Piatek, Pryor, Olszewski, Harris, Mateo,
  Minniti  \& Tinney}{Piatek et~al.}{2003}]{Piatek2003}
Piatek S.,  Pryor C.,  Olszewski E.~W.,  Harris H.~C.,  Mateo M.,  Minniti D.,
   Tinney C.~G.,  2003, \mn@doi [\aj] {10.1086/378713}, 126, 2346

\bibitem[\protect\citeauthoryear{Reid \& Brunthaler}{Reid \&
  Brunthaler}{2004}]{Reid2004}
Reid M.~J.,  Brunthaler A.,  2004, \mn@doi [\apj] {10.1086/424960}, 616, 872

\bibitem[\protect\citeauthoryear{Riello et~al.,}{Riello
  et~al.}{2021}]{Riello2020}
Riello M.,  et~al., 2021, \mn@doi [A\&A] {10.1051/0004-6361/202039587}, 649, A3

\bibitem[\protect\citeauthoryear{Roderick, Mackey, Jerjen  \&
  Da Costa}{Roderick et~al.}{2016}]{Roderick2016}
Roderick T.~A.,  Mackey A.~D.,  Jerjen H.,   Da Costa G.~S.,  2016, \mn@doi
  [\mnras] {10.1093/mnras/stw1541}, 461, 3702

\bibitem[\protect\citeauthoryear{Simon}{Simon}{2018}]{Simon2018}
Simon J.~D.,  2018, \mn@doi [\apj] {10.3847/1538-4357/aacdfb}, 863, 89

\bibitem[\protect\citeauthoryear{Simon}{Simon}{2019}]{Simon2019}
Simon J.~D.,  2019, \mn@doi [\araa] {10.1146/annurev-astro-091918-104453}, 57,
  375

\bibitem[\protect\citeauthoryear{Sohn, Besla, van~der Marel, Boylan-Kolchin,
  Majewski  \& Bullock}{Sohn et~al.}{2013}]{Sohn2013}
Sohn S.~T.,  Besla G.,  van~der Marel R.~P.,  Boylan-Kolchin M.,  Majewski
  S.~R.,   Bullock J.~S.,  2013, \mn@doi [\apj] {10.1088/0004-637x/768/2/139},
  768, 139

\bibitem[\protect\citeauthoryear{Sohn et~al.,}{Sohn et~al.}{2017}]{Sohn2017}
Sohn S.~T.,  et~al., 2017, \mn@doi [\apj] {10.3847/1538-4357/aa917b}, 849, 93

\bibitem[\protect\citeauthoryear{Strigari}{Strigari}{2010}]{Strigari2010}
Strigari L.~E.,  2010, \mn@doi [Advances in Astronomy] {10.1155/2010/407394},
  2010, 407394

\bibitem[\protect\citeauthoryear{Tolstoy et~al.,}{Tolstoy
  et~al.}{2004}]{Tolstoy2004}
Tolstoy E.,  et~al., 2004, \mn@doi [\apj] {10.1086/427388}, 617, L119

\bibitem[\protect\citeauthoryear{Walker}{Walker}{2013}]{Walker2012}
Walker M.,  2013, Dark Matter in the Galactic Dwarf Spheroidal Satellites.
Springer Netherlands, Dordrecht, pp 1039--1089,
  \mn@doi{10.1007/978-94-007-5612-0_20}

\bibitem[\protect\citeauthoryear{Walker, Mateo  \& Olszewski}{Walker
  et~al.}{2008}]{Walker2008}
Walker M.~G.,  Mateo M.,   Olszewski E.~W.,  2008, \mn@doi [\apj]
  {10.1086/595586}, 688, L75

\bibitem[\protect\citeauthoryear{Walker, Mateo  \& Olszewski}{Walker
  et~al.}{2009a}]{Walker2009}
Walker M.~G.,  Mateo M.,   Olszewski E.~W.,  2009a, \mn@doi [\aj]
  {10.1088/0004-6256/137/2/3100}, 137, 3100

\bibitem[\protect\citeauthoryear{{Walker}, {Mateo}, {Olszewski},
  {Pe{\~n}arrubia}, {Evans}  \& {Gilmore}}{{Walker}
  et~al.}{2009b}]{Walker2009sig}
{Walker} M.~G.,  {Mateo} M.,  {Olszewski} E.~W.,  {Pe{\~n}arrubia} J.,  {Evans}
  N.~W.,   {Gilmore} G.,  2009b, \mn@doi [\aj] {10.1088/0004-637X/704/2/1274},
  \href {https://ui.adsabs.harvard.edu/abs/2009ApJ...704.1274W} {704, 1274}

\bibitem[\protect\citeauthoryear{Walker, Olszewski  \& Mateo}{Walker
  et~al.}{2015}]{Walker2015}
Walker M.~G.,  Olszewski E.~W.,   Mateo M.,  2015, \mn@doi [\mnras]
  {10.1093/mnras/stv099}, 448, 2717

\bibitem[\protect\citeauthoryear{Wang et~al.,}{Wang et~al.}{2019}]{Wang2019}
Wang M.~Y.,  et~al., 2019, \mn@doi [\apj] {10.3847/1538-4357/ab31a9}, 881, 118

\bibitem[\protect\citeauthoryear{Wheeler et~al.,}{Wheeler
  et~al.}{2017}]{Wheeler2017}
Wheeler C.,  et~al., 2017, \mn@doi [\mnras] {10.1093/mnras/stw2583}, 465, 2420

\bibitem[\protect\citeauthoryear{White \& Rees}{White \&
  Rees}{1978}]{WhiteRees1978}
White S. D.~M.,  Rees M.~J.,  1978, \mn@doi [\mnras] {10.1093/mnras/183.3.341},
  183, 341

\bibitem[\protect\citeauthoryear{Wilkinson, Kleyna, Evans, Gilmore, Irwin  \&
  Grebel}{Wilkinson et~al.}{2004}]{Wilkinson2004}
Wilkinson M.~I.,  Kleyna J.~T.,  Evans N.~W.,  Gilmore G.~F.,  Irwin M.~J.,
  Grebel E.~K.,  2004, \mn@doi [\apj] {10.1086/423619}, 611, L21

\bibitem[\protect\citeauthoryear{Zhu, van~de Ven, Watkins  \& Posti}{Zhu
  et~al.}{2016}]{Zhu2016}
Zhu L.,  van~de Ven G.,  Watkins L.~L.,   Posti L.,  2016, \mn@doi [\mnras]
  {10.1093/mnras/stw2081}, 463, 1117

\bibitem[\protect\citeauthoryear{de Boer et~al.,}{de~Boer
  et~al.}{2011}]{deBoer2011}
de Boer T. J.~L.,  et~al., 2011, \mn@doi [A\&A] {10.1051/0004-6361/201016398},
  528, A119

\bibitem[\protect\citeauthoryear{de Boer, Tolstoy, Saha  \& Olszewski}{de~Boer
  et~al.}{2013}]{deBoer2013}
de Boer T. J.~L.,  Tolstoy E.,  Saha A.,   Olszewski E.~W.,  2013, \mn@doi
  [A\&A] {10.1051/0004-6361/201220855}, 551, A103

\bibitem[\protect\citeauthoryear{del Pino, Hidalgo, Aparicio, Gallart, Carrera,
  Monelli, Buonanno  \& Marconi}{del Pino et~al.}{2013}]{DelPino2013}
del Pino A.,  Hidalgo S.~L.,  Aparicio A.,  Gallart C.,  Carrera R.,  Monelli
  M.,  Buonanno R.,   Marconi G.,  2013, \mn@doi [\mnras]
  {10.1093/mnras/stt833}, 433, 1505

\bibitem[\protect\citeauthoryear{{del Pino}, Aparicio  \& Hidalgo}{{del Pino}
  et~al.}{2015}]{DelPino2015}
{del Pino} A.,  Aparicio A.,   Hidalgo S.~L.,  2015, \mn@doi [\mnras]
  {10.1093/mnras/stv2174}, 454, 3996

\bibitem[\protect\citeauthoryear{del Pino, Aparicio, Hidalgo  \& {\L}okas}{del
  Pino et~al.}{2017}]{delPino2017}
del Pino A.,  Aparicio A.,  Hidalgo S.~L.,   {\L}okas E.~L.,  2017, \mn@doi
  [\mnras] {10.1093/mnras/stw3016}, 465, 3708

\bibitem[\protect\citeauthoryear{del Pino, Fardal, van~der Marel, {\L}okas,
  Mateu  \& Sohn}{del Pino et~al.}{2021}]{delPino2021}
del Pino A.,  Fardal M.~A.,  van~der Marel R.~P.,  {\L}okas E.~L.,  Mateu C.,
  Sohn S.~T.,  2021, \mn@doi [\apj] {10.3847/1538-4357/abd5bf}, 908, 244

\bibitem[\protect\citeauthoryear{van~der Marel \& Cioni}{van~der Marel \&
  Cioni}{2001}]{VanderMarel2001}
van~der Marel R.~P.,  Cioni M.-R.~L.,  2001, \mn@doi [\aj] {10.1086/323099},
  122, 1807

\bibitem[\protect\citeauthoryear{van~der Marel, Alves, Hardy  \&
  Suntzeff}{van~der Marel et~al.}{2002}]{VanderMarel2002}
van~der Marel R.~P.,  Alves D.~R.,  Hardy E.,   Suntzeff N.~B.,  2002, \mn@doi
  [\aj] {10.1086/343775}, 124, 2639

\bibitem[\protect\citeauthoryear{van~der Marel, Fardal, Sohn, Patel, Besla, del
  Pino, Sahlmann  \& Watkins}{van~der Marel et~al.}{2019}]{VanderMarel2019}
van~der Marel R.~P.,  Fardal M.~A.,  Sohn S.~T.,  Patel E.,  Besla G.,  del
  Pino A.,  Sahlmann J.,   Watkins L.~L.,  2019, \mn@doi [\apj]
  {10.3847/1538-4357/ab001b}, 872, 24

\makeatother
\end{thebibliography}

% Don't change these lines
\bsp	% typesetting comment
\label{lastpage}
\end{document}